\documentclass{aa}
\usepackage[varg]{txfonts}
\usepackage{graphicx}
\usepackage{txfonts}
\usepackage{natbib}
\usepackage{verbatim}
\usepackage[pdftex]{color}
\usepackage[dvipsnames]{xcolor}
\usepackage{hyperref}
\usepackage{ulem}
\usepackage{subfigure}
\hypersetup{
    colorlinks = true,
    citecolor ={blue}
}
\usepackage{mathrsfs}
\bibpunct{(}{)}{;}{a}{}{,} 
\def\msun{M$_\odot$}

\newcommand{\Ion}[2]{#1\,\textsc{#2}}
\usepackage{booktabs}
\definecolor{ultramarine}{rgb}{0.07, 0.1, 0.6} 
\definecolor{myblue}{rgb}{0.07, 0.2, 0.6} 
\definecolor{dopal}{rgb}{.70, .25, .05}

\begin{document}

\title{A hidden population of white dwarfs with atmospheric carbon traces in the Gaia bifurcation}

\author{Maria Camisassa\inst{1},  Santiago Torres\inst{1,2}, Mark Hollands \inst{3}, Detlev Koester \inst{4}, Roberto Raddi\inst{1,2} , Leandro G. Althaus\inst{5,6}, Alberto Rebassa-Mansergas\inst{1,2}}
\institute{Departament de F\'\i sica, 
           Universitat Polit\`ecnica de Catalunya, 
           c/Esteve Terrades 5, 
           08860 Castelldefels, 
           Spain
           \and
           Institute for Space Studies of Catalonia, 
           c/Gran Capit\`a 2--4, 
           Edif. Nexus 104, 
           08034 Barcelona, 
           Spain
           \and
 Department of Physics and Astronomy, University of Sheffield, Sheffield, S3 7RH, UK
 \and 
  Institut für Theoretische Physik und Astrophysik, Christian-Albrechts-Universität, Kiel 24118, Germany
  \and
Grupo de Evoluci\'on Estelar y Pulsaciones. 
           Facultad de Ciencias Astron\'omicas y Geof\'{\i}sicas, 
           Universidad Nacional de La Plata, 
           Paseo del Bosque s/n, 1900 
           La Plata, 
           Argentina
           \and
Instituto de Astrof\'sica de La Plata, UNLP-CONICET, Paseo del Bosque s/n, 1900 La Plata, Argentina   }
\date{Received ; accepted }
\abstract{The high-quality photometric and astrometric capabilities of the ESA {\it Gaia} space mission have revealed a bifurcation of the white dwarf sequence on the color magnitude diagram with two branches: A and B. While the A branch consists mostly of white dwarfs
with hydrogen(H)-rich atmospheres, the B branch is not completely understood. Although invoked to be populated mainly by helium(He)-rich white dwarfs, the B branch overlaps a $\rm \sim 0.8$\,M$_\odot$ evolutionary track with a pure He envelope, fact that would imply an unexpected peak in the white dwarf mass distribution. 
}   
{
In cold He-rich white dwarfs, it is expected that the outer convective zone penetrates into deep carbon(C)-rich layers, thus leading to a slight C contamination  in their surfaces at $\sim 10\,000$\,K. In this paper we aim at studying the {\it Gaia} bifurcation as the natural consequence of C dredge-up by convection in cold He-dominated white dwarfs.}
{
Relying on accurate atmosphere models, we provide a new set of evolutionary models for He-rich white dwarfs employing different prescriptions for the C enrichment. On the basis of these models, we made a population synthesis study of the {\it Gaia} 100\,pc white dwarf sample to constrain the models that best fit the bifurcation.}
{
Our study shows that He-rich white dwarf models with a slight C contamination below the optical detection limit can accurately reproduce the {\it Gaia} bifurcation. We refer to these stars as ``stealth DQ'' white dwarfs because they do not exhibit detectable C signatures in their optical spectra, but the presence of C in their atmosphere produces a continuum absorption favouring the emission in bluer wavelengths, thereby creating the B branch of the bifurcation. Furthermore, our study shows that the white dwarf mass distribution obtained when a stealth C contamination is taken into account presents a peak at $\rm \sim 0.6$\,M$_\odot$, which is consistent with the mass distribution for H-rich white dwarfs and with the standard evolutionary channels for their formation.
 }
{
We conclude that ``stealth DQ'' white dwarfs can account for the lower branch in the {\it Gaia} bifurcation. The C signatures of these stars could be detectable in Ultra-Violet (UV) spectra.}

\titlerunning{A hidden population of white dwarfs with trace carbon}
\authorrunning{Camisassa et al.}

\maketitle

\section{Introduction}
\label{introduction}

 White dwarfs are the  most common end point of stellar evolution, as they are the final destiny of more than 95\% of the main sequence stars. These old compact objects, supported by electron-degeneracy pressure, undergo a slow cooling process that lasts for several Gyrs, turning these objects into reliable cosmochronometers to date stellar populations and main sequence companions ---  see  for  instance, the  reviews  of
\cite{2008PASP..120.1043F},       \cite{2008ARA&A..46..157W},      
\cite{2010A&ARv..18..471A}, \cite{2016NewAR..72....1G} and \cite{2019A&ARv..27....7C}. Due to their unique characteristics, white dwarf stars are considered important objects for understanding the late stages of stellar evolution, as well as planetary systems and the structure and evolution of our Galaxy. Also, they can be used to infer the star formation rate, the initial mass function, the initial to final mass relation and the chemical evolution in the solar neighborhood \citep[e.g.][]{2021MNRAS.505.3165R,2022A&A...658A..22R}. Furthermore, the extreme densities that characterize the white dwarf interior, turn these stars into promising laboratories to study stellar matter and energy sources under extreme conditions \citep[e.g.][]{2022FrASS...9....6I}. 

In the last decades, we have entered a golden era for the exploitation of white dwarf science. Surveys like Sloan Digital Sky Survey \citep[SDSS,][]{2000AJ....120.1579Y}, the Radial Velocity Experiment \citep[RAVE, ][]{Steinmetz_2020II,Steinmetz_2020}, the Panoramic Survey Telescope and Rapid Response System \cite[PanSTARRS,][]{2016arXiv161205560C}
and others, are providing the first large samples of moderate-resolution spectra and multi-band photometry for stars in
our Galaxy \citep[e.g.][]{Kepler2021}, and missions like NASA Kepler and NASA  Transiting Exoplanet Survey Satellite \citep[TESS, ][]{2015JATIS...1a4003R} are providing measurements of photometric variations of these stars. 
In particular, the successive data releases by the ESA space mission {\it Gaia} constitute an unprecedented advance, providing multi-band photometry, synthetic spectra, proper motions and parallaxes for more than a billion sources \citep{GaiaDR22018,GaiaEDR32021}. Among these, \cite{Fusillo2021} identified nearly  360\,000 high-confident white dwarf candidates in the {\it Gaia} Data Release 3 (DR3), of which nearly 13\,000 are within the 100 pc volume-limited sample \citep{Jimenez2018,2023MNRAS.518.5106J}, leading the white dwarf research field into a new era.

The power of {\it Gaia} space mission has revealed some unexpected features in the white dwarf cooling sequence on the color magnitude diagram, identified as the A, B and Q branches \citep[see][]{GaiaDR22018}.  The A-branch is mainly populated by white dwarfs with H-rich atmospheres, that is DA spectral type, and overlaps with the evolutionary track of an approximately 0.6\,M$_\odot$ white dwarf. On the other hand, the B-branch constitutes a bifurcation from the A-branch, has a significant fraction of He-rich white dwarfs, and lies on a $\sim$0.8\,M$_\odot$ evolutionary track with a pure He envelope. Finally, the Q branch has a weaker concentration of white dwarfs and does not follow any evolutionary track nor isochrone. 
 
The origin of the Q branch has been extensively discussed in the literature, with the general consensus being that it arises from an energy released during white dwarf core crystallization \citep{2019Natur.565..202T,Cheng2019,2021A&A...649L...7C,2021ApJ...911L...5B,2021ApJ...919L..12C,2022MNRAS.511.5198C,2020ApJ...902...93B,2022MNRAS.511.5984F}. In contrast, the origin of the AB bifurcation remains unclear, with no general consensus. \cite{BadryIFMR} attributed the existence of the bifurcation to a flattening in the initial-to-final-mass relation (IFMR), which leads to a secondary peak in the white dwarf mass distribution at approximately $0.8$\,M$_\odot$. 
Similarly, \cite{2018MNRAS.479L.113K} also suggested the presence of this secondary peak, but they attributed it to the occurrence of stellar mergers. Alternatively, \cite{2020MNRAS.492.5003O} proposed a different explanation for the origin of the {\it Gaia} bifurcation, suggesting that spectral evolution from a pure H to pure He envelope at an effective temperature ($T_\mathrm{eff}$) $\sim 10\,000$\,K in approximately 16\% of DA white dwarfs may be an important contributing factor.
Nevertheless, a recent study by \cite{Bergeron2019} found that assuming a pure He atmosphere for all non-DA white dwarfs leads to a low number of objects with masses around $\sim0.6$\,M$_\odot$ when $T_\mathrm{eff}<11\,000$\,K,
which is inconsistent with the observed mass distribution at higher effective temperatures. These authors, therefore, suggested to consider a He atmosphere with traces of H instead \citep[see also][]{2019A&A...623A.177S}. The additional electrons provided by traces of H in a He-rich white dwarf atmosphere cause a shift in the 0.6\,M$_\odot$ evolutionary track, when $T_\mathrm{eff}<11,000$\,K , thus creating the bifurcation. However, \cite{Bergeron2019} noticed that a significant fraction of He-rich white dwarfs should have H abundances low enough for not to affect their photometric mass determinations
at $\rm 8\,000\,\mathrm{K}\lesssim T_\mathrm{eff} \lesssim 10\,000$\,K, and, therefore, still a significant number of pure He white dwarfs with $\sim 0.6$\,M$_\odot$ would be expected. Thus, these authors suggested that another electron donor, such as C or metals, is required to fully explain the origin of the {\it Gaia} bifurcation.

In this paper we aim at studying the B branch as a result of the presence of C in the atmosphere of He-dominated white dwarfs. Theoretical models predict that He-dominated white dwarfs will dredge-up C as a result of convective mixing in the so called PG1159-DO-DB-DQ Spectral Evolutionary Channel \citep[see] []{1982A&A...116..147K,1986ApJ...307..242P,2005A&A...435..631A, 2005ApJ...627..404D,Camisassa2017,BedardDQ}. 
Unfortunately, the exact amount of C dredged-up in a He rich white dwarf cannot be predicted by theoretical models. In particular, \cite{BedardDQ} followed the C enrichment from the beginning of the white dwarf cooling phase under different initial conditions and physical inputs, finding that the amount of C dredged-up by convection depends on the initial C surface abundance, the thickness of the He layer, the efficiency of extra-mixing beyond the convective boundaries, and the stellar mass. 
We find that the {\it Gaia} bifurcation can be explained by He rich white dwarfs that experience a spectral evolution at $T_\mathrm{eff}\sim 12\,000$\,K with a smooth C contamination just below the detection limits for optical spectroscopy. While this C contamination does not produce C lines in the optical spectra, it adds free electrons to the He envelope, that produce a continuum opacity that shifts the $0.6$\,M$_{\odot}$ evolutionary track causing the  {\it Gaia} bifurcation. This C contamination, however, has strong features in the UV spectra that could potentially be detectable. 

This paper is organized as follows.
In Sect. \ref{wdmodels} we describe the
white dwarf evolutionary models employed. In Sect. \ref{carbonenrichment} we describe the C enrichment observed in white dwarfs determined in previous studies and the different prescriptions for the C enrichment adopted in this paper. In Sect. \ref{atmospheres} and \ref{MCcode} we present details on the atmosphere models and the population synthesis code employed, respectively. In Sect. \ref{Results} we describe our results and, finally, in Sect. \ref{conclusions}  we summarize the main findings of the paper.

\section{Methods}

\subsection{The white dwarf cooling models}
\label{wdmodels}

White dwarfs can be categorized based on the presence or the absence of H in their atmospheres.  It is thought that H is dominant in the atmosphere of nearly $80\%$ per cent of all white dwarfs (the so-called DA white dwarf class), and that the remaining $20\%$ are depleted in H (the so-called non-DA white dwarf class) \citep{Kepler2021,2023MNRAS.518.5106J}.  
The formation and evolution of H-rich white dwarfs is reasonably well understood and has been computed in several studies in the literature \citep[for recent studies see for instance][]{2010ApJ...717..183R,2015A&A...576A...9A,Camisassa2016,2019A&A...625A..87C,2020ApJ...901...93B,2022MNRAS.509.5197S}, and their pure H envelopes are the natural consequence of H floating up due to gravitational settling acting on an initial mixed H/He composition. 
Although a fraction of the H-rich white dwarfs could undergo spectral evolution due to convective dilution  \citep{2023arXiv230205424B}, in this paper we assume that H-rich white dwarfs will retain their H envelope through their evolution. 

While DA white dwarfs have H-dominated atmospheres, most of the non-DA white dwarfs have atmospheres that are dominated by He. 
The evolution of white dwarfs with He-dominated atmospheres has been extensively studied during the last decades \citep{2005A&A...435..631A,Camisassa2017,2022MNRAS.509.5197S,BedardDQ}. 
It is thought that He-rich white dwarfs are the descendants of the PG\,1159 stars, hot stars with an envelope composed by He, C, and oxygen (O) in similar amounts. Initially, the C and O in the outer layers are supported by a weak radiative wind but as the white dwarf cools down, the wind weakens and the C and O rapidly sink into the stellar interior as a result of gravitational settling, leading to a pure-He atmosphere. This pure-He enveloped white dwarf should present He absorption lines, being classified first as a DO (\Ion{He}{ii} lines) and later on as a DB (\Ion{He}{i} lines).  
 As the white dwarf cools down, a convection zone gradually develops within the He envelope, growing inward and ultimately reaching the previously settled C, 
which, depending on the He layer thickness,
 is carried back to the surface. This spectral evolution transforms the pure He DB white dwarf into a DQ white dwarf, a He-dominated atmosphere white dwarf containing traces of C.

In this paper, we employ the H-rich evolutionary models of \cite{Camisassa2016}, which are the result of the full evolution of progenitor stars starting at the Zero Age Main Sequence and evolved through the central H and He burning, the Asymptotic Giant Branch (AGB) and the post-AGB phases, as calculated in \cite{2016A&A...588A..25M}. These models commence the white dwarf phase with a H-dominated atmosphere with small amounts of He, C and O. However, due to gravitational settling, the heavier elements rapidly sink, causing the outer envelope of the white dwarf to become entirely composed of H.

For He-dominated white dwarfs (non-DA), we employed the evolutionary models calculated in \cite{Camisassa2017}, which are the result of the full progenitor evolution through the born again scenario \citep[see][for details]{Camisassa2017,2006A&A...454..845M}. These models follow the complete white dwarf evolution from the PG\,1159 stage at the beginning of the white dwarf cooling sequence, all the way to very low luminosities, keeping track  of 
the gravitational settling of  the C left in the envelope by the born again evolution, and its later convective  dredge-up.
Although these models do not predict the correct amount of C dredge-up to the surface, we amended this issue by employing different artificial parametrizations for the C abundance that match the observed C sequence (see Sect. \ref{carbonenrichment}).

It is worth noting that the cooling times considered in these models are expected to be realistic, as they take into account all relevant sources and sinks of energy and use accurate initial chemical profiles derived from the full progenitor evolution.

\begin{figure}
        \centering
        \includegraphics[width=1.\columnwidth]{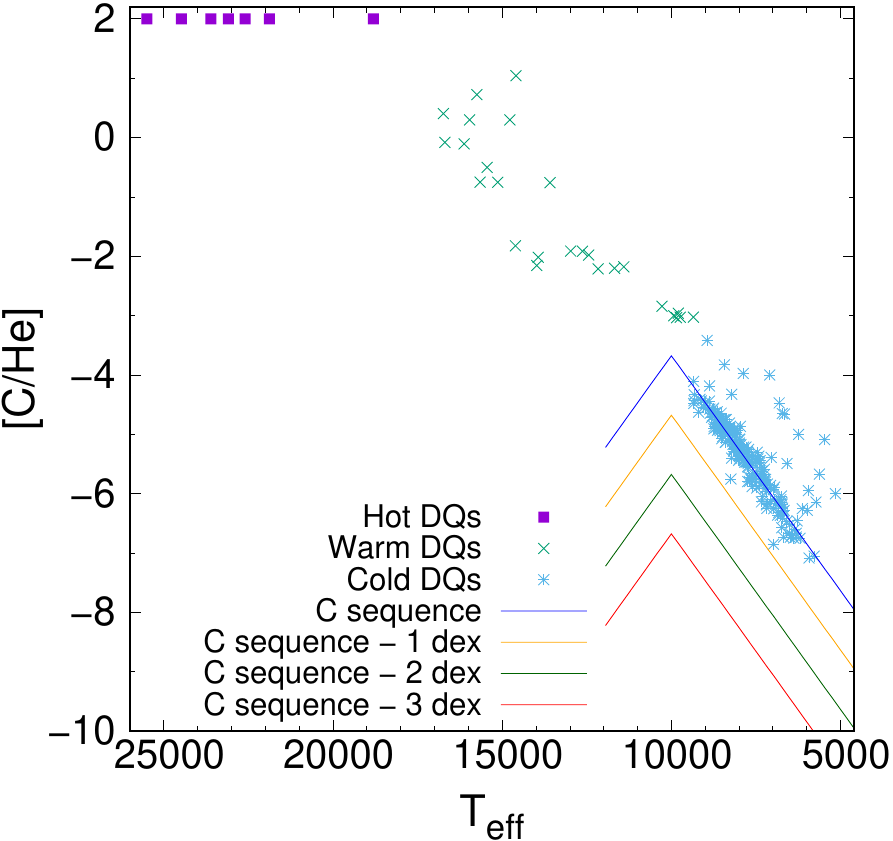}
        \caption{Carbon to helium surface abundance ratio ([C/He]) vs. effective temperature for the hot DQs (purple squares), warm DQs (green crosses), and cold DQs (light-blue asterisks) taken from \cite{2019A&A...628A.102K}. We note that, for the hot DQs, [C/He] has been fixed as the typical lower limit. The blue line is our parametrization for the observed [C/He] trend (the C sequence), and the orange, green and red lines are the C sequence -1, -2 and -3 dex, respectively (see text for details). 
       } 
        \label{cparam}
\end{figure}

\subsection{The carbon enrichment}
\label{carbonenrichment}

As we mentioned, theoretical models suggest that He-rich white dwarfs experience spectral evolution when their outer convective zone penetrates into C-rich layers, transforming them into DQ white dwarfs. The surface C abundance should grow until the convective zone reaches its maximum depth, and then it starts to decrease gradually as a result of C  recombination at the bottom of the convective zone, which makes C sink into the interior \citep{1986ApJ...307..242P}. A recent research conducted by \cite{BedardDQ} investigated the evolution of the surface C abundance under different assumptions and found that it is strongly influenced by the initial 
conditions and physical inputs considered in the modeling. First, a higher C abundance at the beginning of the white dwarf phase, as well as a thinner He envelope, would be reflected in a larger amount of C dredged-up by convection \citep[see Figs. 9 and 10 of][respectively]{BedardDQ}. In addition, the stellar mass and the efficiency of extra-mixing beyond the convective boundaries play a key role in determining the C enrichment sequence. Thus, while theoretical models predict convective C dredge-up and a subsequent decrease in C abundance due to its recombination, they cannot accurately determine the exact C abundance in cold non-DA white dwarfs. Therefore, observational data is necessary to provide insights into this matter.

\begin{figure}
        \centering
        \includegraphics[width=1.\columnwidth]{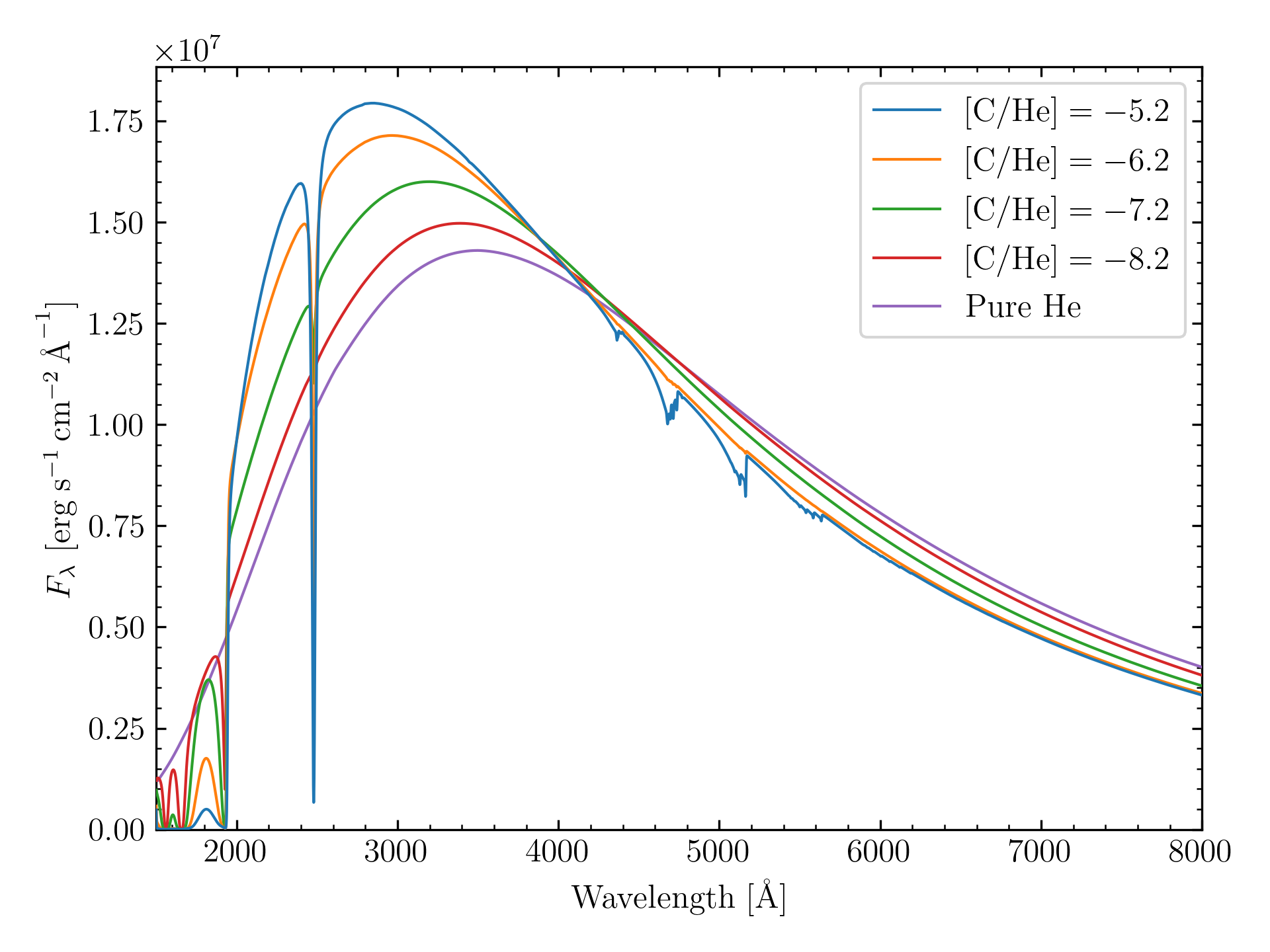}
        \caption{Synthetic spectra for atmosphere models with $\log g=8.0$ and $T_\mathrm{eff}=8\,000$\,K. The pure He model is displayed using a purple line. The other models assume [C/He]= -5.2 (blue line, the C sequence), [C/He]= -6.2 (orange line, the C sequence - 1 dex), [C/He]= -7.2 (green line, the C sequence -2 dex), [C/He]= -8.2 (red line, the C sequence -3 dex).} 
        \label{spectra}
\end{figure}

The white dwarfs showing C lines or molecular C in their spectra can be classified into three groups with markedly different characteristics: hot, warm and cold DQs. In Fig.~\ref{cparam} we plot the logarithm of the ratio between the numerical abundances of C and He ([C/He]) vs. effective temperature determinations from  \cite{2019A&A...628A.102K} for the hot, warm and cold DQs, using purple squares, green crosses and light-blue asterisks, respectively. 
It should be noted that [C/He] for hot DQs has been set to the typical lower limit. The hot and warm DQs have considerably higher velocities, masses and C abundances than cold DQs. 
Therefore, it is very unlikely that the hot and warm DQs are the predecessors of cold DQs, and their origin has been attributed to stellar mergers \citep{2015ASPC..493..547D,2023MNRAS.520.6299K}. Furthermore, theoretical models of convective C dredge-up in He-rich envelopes fail to reproduce C enrichment in hot and warm DQs, but they can account for the C abundances at $T_\mathrm{eff}\lesssim 12\,000\,\mathrm{K}$ observed in cold DQs.
In a nutshell, cold DQs are thought to be the descendants of He-rich DB white dwarfs, that became C enriched when the outer convective zone penetrated into C rich layers. 

\cite{2020A&A...635A.103K} estimated the total He mass in the envelope of cold DQs, by integrating the envelope equations from the outside, using observed surface abundances as starting point. 
These authors found that  the total He mass fraction, q(He), in the envelope of cold DQs is independent of their effective temperature. They also found that the convection zone is marginally deeper for the colder DQs. These results support the idea that the cold DQ white dwarfs constitute an evolutionary path of white dwarfs with similar characteristics, and that the gradual decrease in [C/He] is caused by the C recombination at the base of the He envelope. Even when including overshooting in their calculations, these authors obtained q(He)$\sim -3.$, which is nearly one order of magnitude lower than predicted from stellar evolution \citep{2006A&A...454..845M,2009ApJ...704.1605A}. 
Therefore, we can speculate that white dwarfs may have a range of He masses wider than expected, possibly depending on the number of thermal pulses experienced in the progenitor evolution, and that the cool DQs are originated from the white dwarfs with the thinnest He envelopes.

In this paper, we present He-rich white dwarf models (non-DA) that follow the PG1159-DO-DB-DQ evolutionary connection, and experience C dredge-up as a result of  convective dilution. In order to mimic the C enrichment sequence, we consider that these white dwarfs have a pure He envelope if $T_\mathrm{eff}>12\,000\,\mathrm{K}$ 
and a He envelope with C traces if  $T_\mathrm{eff}<12\,000$\,K. We considered four different prescriptions for the [C/He] ratio when $T_\mathrm{eff}<12\,000$\,K. The first set of white dwarf models follows the observed C enrichment sequence in terms of the effective temperature (which we will call ``C sequence'' from now on). The C sequence follows a linear least square fit to the observed C sequence reported in Table~1 of \cite{2020A&A...635A.103K} when $T_\mathrm{eff}<10\,000$\,K. To parameterize the range $\rm 12\,000\,\mathrm{K}>T_\mathrm{eff}>10\,000$\,K, we  reflected this linear fit from $\rm T_\mathrm{eff}=10\,000$\,K, mimicking the rise in the C abundance due to the deepening of the outer convective zone.
The run of [C/He] in terms of the effective temperature of the C sequence is
depicted using a blue line in Fig.~\ref{cparam}. The cool region ($T_\mathrm{eff}\lesssim 10\,000$\,K) of the C sequence is approximately on the optical detection limit of C for a signal-to-noise ratio S/N = 20 according to the DQ model atmospheres of \cite{2019ApJ...878...63B}. 
The [C/He] ratio in terms of the effective temperature in the second, third and fourth sets of white dwarf models is that of the C sequence, but shifted -1 dex , -2 dex and -3 dex, respectively (orange, green and red lines in Fig.~\ref{cparam}). Although C would still be present in their atmospheres, these three sets of models would not show C features in their optical spectra, thus being classified as DC (see Sect. \ref{atmospheres}). These models account for the white dwarfs with thicker He envelopes than the one obtained for the C sequence in \cite{{2020A&A...635A.103K}}.
When [C/He] reaches values lower than $-10.41$, we  considered a pure He atmosphere. 
It is important to remark that we only implemented this C enrichment prescription in the atmosphere models, but the white dwarf structure and evolution is that of \cite{Camisassa2017}, where the C enrichment as a result of convective dredge-up is considered, but it does not follow the enrichment that we desire to simulate.

\subsection{The atmosphere models}
\label{atmospheres}

We calculated a grid of model atmospheres from \cite{2010MmSAI..81..921K} and \cite{2019A&A...628A.102K} for different $T_\mathrm{eff}$, surface gravity and chemical compositions, and then integrated their fluxes in different passbands. In particular, we employed three sets of atmosphere models: one with a pure H composition, one with a pure He composition and one with a mixed He/C composition. The He/C ratio in the mixed composition models can vary according the C enrichment prescription described in Sect. \ref{carbonenrichment}.

In Fig.~\ref{spectra} we show the synthetic spectra for atmosphere models with $\log g = 8.0$ and $T_\mathrm{eff} = 8\,000\,$\,K and different chemical compositions: [C/He]=-5.2 (C sequence),
[C/He]= -6.2 (C sequence - 1 dex), [C/He]= -7.2 (C sequence -2 dex), [C/He]= -8.2 (C sequence -3 dex), and pure He. The spectrum of the C sequence atmosphere has strong C features, both in the optical and UV wavelengths. On the contrary, the atmosphere models with lower [C/He] and the pure He model do not show spectral lines nor molecular Swan bands in the optical, resembling a continuum spectrum, which would be classified as a DC if observed at these wavelengths. Regarding the UV spectra, the C-contaminated models do show \Ion{C}{i} absorption lines at 1931\,\AA\ and 2479\,\AA, which are more noticeable when the C abundance is higher.  We decided to call the cold white dwarfs ($T_\mathrm{eff}\lesssim 12\,000$\,K) with a trace [C/He] abundance lower than the C sequence ``stealth DQs'', which do not present C features in the optical spectra but do show them in the UV wavelengths.

Although the ``stealth DQ'' white dwarfs would appear as DC when observed in the optical wavelengths, their optical fluxes still differ from a pure He model, since their continuum emission is shifted to bluer wavelengths (see Fig.~\ref{spectra}). The more abundant the trace C is, the bluer the white dwarf becomes. The origin in this shift is 
caused by   an increase in the He$^-$ free-free absorption, which is markedly altered by the presence of C. At low effective temperatures, only a tiny fraction of He becomes ionised to provide free electrons, and so only a very small amount of He$^-$ can form in a pure He atmosphere. 
Because C has a much lower ionisation potential than He,
only trace amounts of C are required for it to become the
primary electron donor
allowing for a higher number of He$^-$ ions to form, which in turn increases the He$^-$ opacity, causing the shift to bluer wavelengths.

\subsection{The population synthesis code}
\label{MCcode}

We performed a population synthesis analysis of the white dwarf thin disk population within 100 pc, previously classified in \cite{Torres2019} to provide insights on the C enrichment sequence. We employed a Monte Carlo population synthesis code widely used in the study of the single \citep[e.g.][]{GBerro1999,Torres2005,Torres2016,Jimenez2018,Torres2021} and binary \citep[e.g.][]{Camacho2014,Cojocaru2017,2018MNRAS.480.4519C,2022MNRAS.511.5462T} white dwarf population, as well as on   studies of open and globular clusters \citep[e.g][]{2010Natur.465..194G,Torres2015} and the Galactic bulge \citep{Torres2018}. 
A detailed description of the code can be
found in these references. 
Therefore, in this paper, we will provide a brief overview of its key inputs.
 
 Synthetic main sequence stars are generated with masses randomly following an initial mass function with a Salpeter distribution, considering $\alpha= -2.35$ and a minimum mass of  0.4\,\msun. We assume a constant star formation rate, with a maximum age of  10.5 Gyrs. Alternative prescriptions for the star formation rate and the total age will not affect the robustness of our results, since we are interested in analyzing only the  {\it Gaia} bifurcation. Once each main sequence star is generated, we employ the pre-white dwarf age of the Basti database \citep{bastinew} to see which stars had time to become white dwarfs. Then, using an IFMR we can obtain the white dwarf masses and cooling times. We considered two different IFMRs: \cite{CatalanIFMR} and \cite{BadryIFMR}. The initial metallicity of all the synthetic stars in this study was fixed to $Z=0.02$, since we do not expect the metallicity distribution to alter our results. 
 Once knowing the white dwarf cooling time and mass, we employed the white dwarf evolutionary models of the La Plata group described in Sect. \ref{wdmodels}, to obtain its physical properties, such as luminosity, effective temperature, surface gravity and radius. We randomly assign an envelope composition to each white dwarf, either H-dominated (DA) or He-dominated (non-DA). We considered four different proportions of DA to non-DA white dwarfs: 100:0, 80:20, 75:25, and 70:30, respectively. For DA white dwarfs, we assumed that they preserve a pure H envelope through their evolution. For non-DA white dwarfs, we consider a pure He envelope if $ T_\mathrm{eff}>12\,000$\,K and, if $T_\mathrm{eff}<12\,000$\,K, we can consider that either the white dwarf retains this pure He envelope or that it undergoes C enrichment. In the simulations where we consider C enrichment in the synthetic populations, we either assume the C sequence -0 dex, -1 dex, or -2 dex. We  also simulated the possibility that each white dwarf may follow a different C enrichment sequence, by randomly assigning a number in the range from 0 to 3 with a uniform distribution and subtract this number to the [C/He] expected for the C sequence. We named this type of C enrichment as ``C random [0:3]'', because it assumes that each individual white dwarf can undergo a different C enrichment sequence. A similar C enrichment sequence was also considered, named ``C random [0:2]'', which assumes a random C distribution, but this time the random number assigned can vary in the range from 0 to 2, thus being this synthetic population more C enriched.
Finally, in order to compare with the observational sample, we employ the atmosphere models described in Sect. \ref{atmospheres} to convert the quantities from our evolutionary models into magnitudes in the {\it Gaia} passbands, and we added observational uncertainties by introducing photometric and astrometric errors in concordance with {\it Gaia} performance \footnote{\url{http://www.cosmos.esa.int/web/gaia/science-performance}}.
  
A total of 22 synthetic white dwarf populations models were generated, varying the proportion of DA to non-DA, the non-DA envelope composition, and the IFMR. The main characteristics of these synthetic population models
are described in Table~\ref{table1}.

\begin{table*}[t]
\centering
\begin{tabular}{|l|c|c|c|} 
  \hline
  \hline
Ratio DA non-DA &  Non-DA composition & IFMR  & $\chi^2$\\
\hline
$100\%$ DA & - & (1)  & 4.046\\
$100\%$ DA & - & (2)  & 8.530 \\
\hline
$80\%$ DA, $20\%$ non-DA & Pure He & (1)  & 3.816 \\
$80\%$ DA, $20\%$ non-DA & Pure He & (2) & 7.863 \\
$80\%$ DA, $20\%$ non-DA & C sequence & (1)  & 2.972 \\
$80\%$ DA, $20\%$ non-DA & C sequence & (2)  & 6.447 \\
$80\%$ DA, $20\%$ non-DA & C sequence -1 dex & (1)  & 3.191 \\
$80\%$ DA, $20\%$ non-DA & C sequence -1 dex & (2)  & 6.801 \\
$80\%$ DA, $20\%$ non-DA & C sequence -2 dex & (1)  & 3.378\\
$80\%$ DA, $20\%$ non-DA & C sequence -2 dex & (2)  & 7.019 \\
$80\%$ DA, $20\%$ non-DA & C random [0:2] & (1)  & 3.042 \\
$80\%$ DA, $20\%$ non-DA & C random [0:3] & (1)  & 3.147 \\
$80\%$ DA, $20\%$ non-DA & C random [0:3] & (2)   & 6.693
 \\
\hline
$75\%$ DA, $25\%$ non-DA & C sequence & (1)   & 3.116 \\
$75\%$ DA, $25\%$ non-DA & C sequence -1 dex & (1)   & 3.228 \\
$75\%$ DA, $25\%$ non-DA & C random [0:2] & (1)   & 3.153 \\
\hline
$70\%$ DA, $30\%$ non-DA & Pure He & (1)  & 4.097 \\
$70\%$ DA, $30\%$ non-DA & C sequence & (1)   &  3.154 \\
$70\%$ DA, $30\%$ non-DA & C sequence -1 dex & (1) &  3.380\\
$70\%$ DA, $30\%$ non-DA & C sequence -2 dex & (1)   & 3.735 \\
$70\%$ DA, $30\%$ non-DA & C random [0:2] & (1) &   3.079 \\
$70\%$ DA, $30\%$ non-DA & C random [0:3] & (1) &    3.126 \\

  \hline  
  
  \end{tabular}                 
  \caption{Main characteristics of each of our synthetic population models and the $\chi^2$ values obtained from the comparison with the observed sample (see text for details).  References: (1) \cite{CatalanIFMR}; (2) \cite{BadryIFMR} }
    \label{table1}
\end{table*}

\begin{figure}
        \centering
        \includegraphics[width=1.\columnwidth]{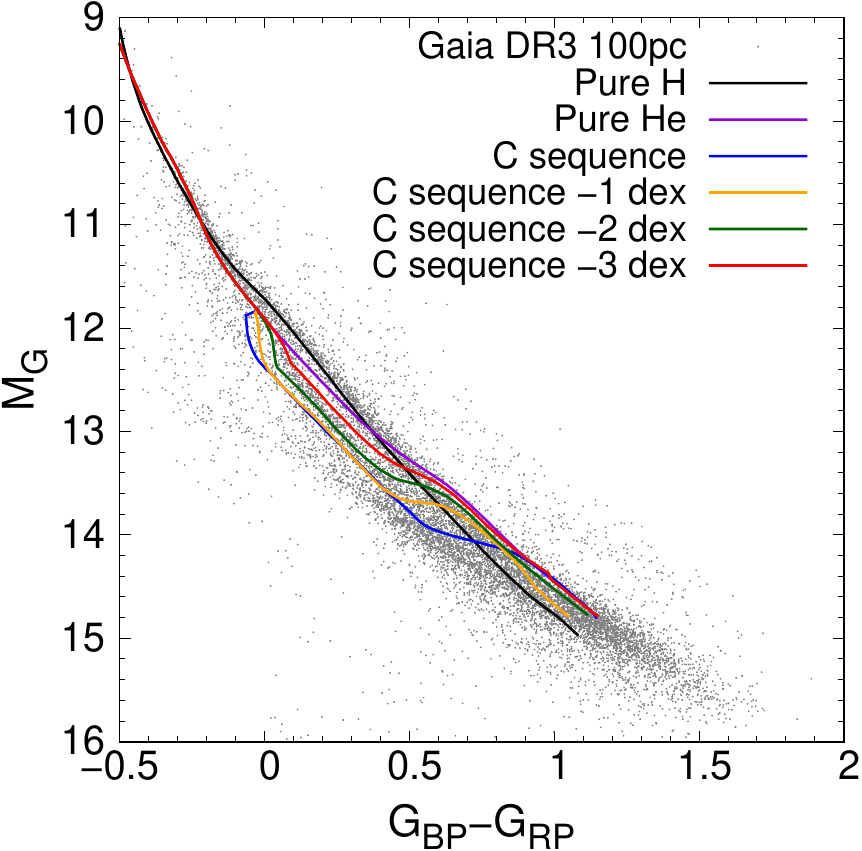}
        \caption{Evolutionary tracks of a         
        0.58\,M$_\odot$ white dwarf for different atmospheric compositions on the {\it Gaia} DR3 color magnitude diagram, together with the observed 100\,pc sample (gray dots).
         The evolutionary model considering a pure H (He) envelope is displayed using a black (purple) line.  Models with a He atmosphere with C traces, considering the C sequence, the C sequence -1 dex, -2 dex and -3 dex are shown using blue, orange, green and red lines, respectively.
        }
        \label{all}
\end{figure}

\section{Results}
\label{Results}

\begin{figure}
        \centering
        \includegraphics[width=1.\columnwidth,trim={20 5 55 30},clip]{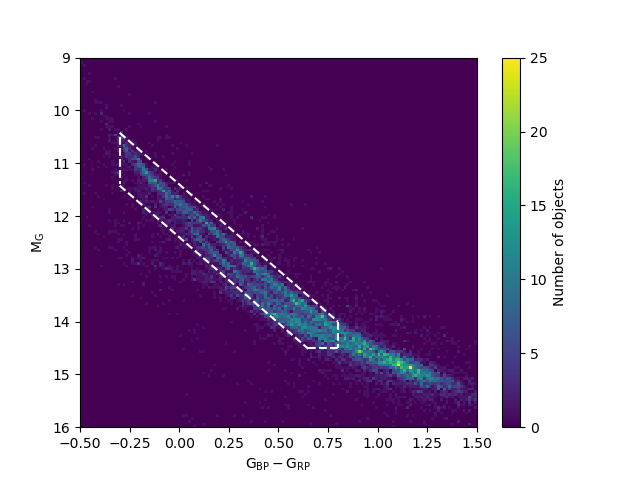}
        \caption{Stellar density (Hess) diagram for the 100\,pc thin disk white dwarf sample in {\it Gaia} DR3. The white dashed lines delimit the region in which we perform the statistical analysis.} 
        \label{hessobs}
\end{figure}

\subsection{Effects of carbon enrichment on the {\it Gaia} color magnitude diagram}

Fig.~\ref{all} displays the effect of the atmospheric composition on the white dwarf evolutionary models in the {\it Gaia} color magnitude diagram. The gray dots are the {\it Gaia} DR3 observations of the white dwarfs within 100\,pc, whereas the solid lines depict 0.58\,M$_\odot$ evolutionary models under different atmospheric compositions. In this plot we can see the {\it Gaia} bifurcation in the A and B branches, starting at $\rm G\sim 12$ and $\rm G_{BP} - G_{RP}\sim 0$. We note that 
that the model with a pure H envelope (black solid line) overlaps with the upper branch of the {\it Gaia} bifurcation. Additionally, we can see that, although a 0.58\,M$_\odot$ pure He model (purple line) can reproduce somewhat better the lower branch of the bifurcation than a 0.58\,M$_\odot$ pure H model, a higher mass ($\sim$0.8\,M$_\odot$) pure He model would overlap with this branch. The blue, orange, green and red lines display 0.58\,M$_\odot$ cooling sequences that mimic the C dredge-up enrichment. These cooling sequences have a pure He atmosphere if $T_\mathrm{eff}>12\,000\,K$ and a He atmosphere with traces of C if $T_\mathrm{eff}<12\,000\,\mathrm{K}$. The C enrichment sequence in each of these evolutionary models is described in Sect. \ref{carbonenrichment}. By inspecting this figure, we can see that the 0.58\,M$_\odot$ models with C enrichment overlap the lower branch of the {\it Gaia} bifurcation, and that the pure H and pure He models do not. As expected, the lower the trace C abundance, the closer to the pure He model appears the evolutionary sequence. We can conclude that the {\it Gaia} bifurcation occurs at $T_\mathrm{eff}\sim 10\,000\,\mathrm{K}$, which is roughly where we expect that the surface C abundance reaches its maximum value, regardless of the input conditions on the modeling \citep[see Figs.~7--12 in][]{BedardDQ}.

It is important to recall that, a C trace abundance just 1 dex below the observed C sequence would be undetectable in optical observations. Therefore, trace C dredge-up by convection in a He dominated atmosphere can be the source of opacity that causes the {\it Gaia} bifurcation. This conjecture is supported by the fact that most of the white dwarfs in the lower branch of the {\it Gaia} bifurcation are non-DA white dwarfs \citep[65\% according to][] {2023MNRAS.518.5106J}.

\subsection{Population synthesis analysis}

\begin{figure}
        \centering
        \includegraphics[width=1.\columnwidth,trim={20 5 30 30},clip]{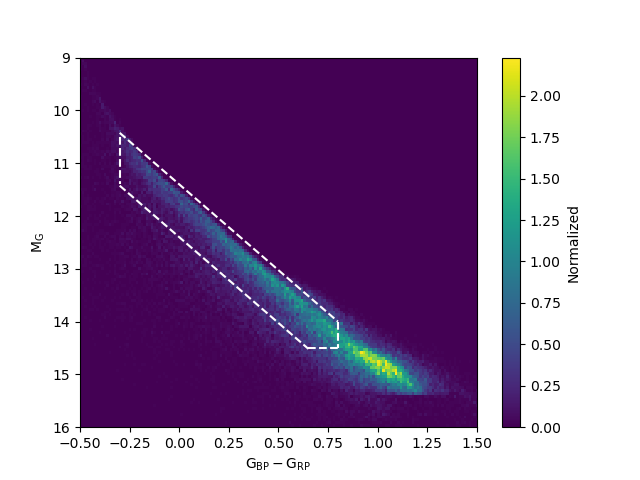}
        \caption{Normalized Hess diagram for a synthetic 100\,pc thin disk white dwarf population
        considering 80\% DA, 20\% non-DA, pure He composition for all non-DA white dwarfs and an IFMR of \cite{CatalanIFMR}.} 
        \label{8020pureHe}
\end{figure}

\begin{figure}
        \centering
        \includegraphics[width=1.\columnwidth,trim={20 5 30 30},clip]{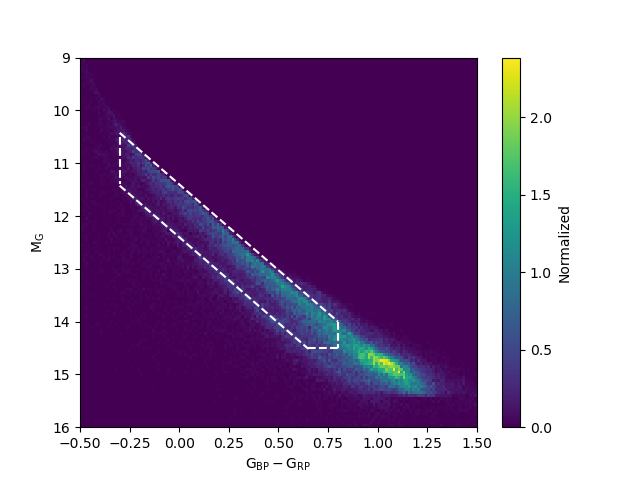}
        \caption{Same as Fig.~\ref{8020pureHe}, but considering an IFMR of \cite{BadryIFMR}.} 
        \label{8020pureHebadry}
\end{figure}

\begin{figure}
        \centering
        \includegraphics[width=1.\columnwidth,trim={20 5 30 30},clip]{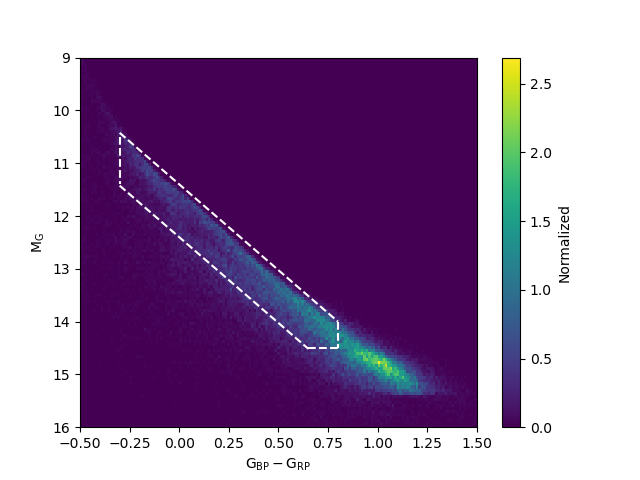}
        \caption{Normalized Hess diagram for a synthetic 100\,pc thin disk white dwarf population
        considering 80\% DA, 20\% non-DA, an IFMR of \cite{CatalanIFMR}, pure He composition for all non-DA white dwarfs with $T_\mathrm{eff}>12\,000 \mathrm K$ and a random C enrichment prescription (C random [0:2]) for all non-DA white dwarfs with $T_\mathrm{eff}<12\,000\, \mathrm K$.} 
        \label{8020Crandom2}
\end{figure}

In order to test the soundness of the C enrichment in He dominated white dwarfs as the mechanism responsible for creating the lower branch of the {\it Gaia} bifurcation, we  performed a population synthesis analysis of the 100\,pc thin disk white dwarf population. The stellar density in the colour–magnitude diagram (Hess diagram) of the observed sample is shown in Fig.~\ref{hessobs}. The {\it Gaia} bifurcation can easily be seen in this stellar density plot. This color-magnitude diagram was divided in $2\,500$ square bins, where we counted the number of stars and then normalized to the total number of observed objects  to perform a $\chi^2$ statistical test 
 \citep{1999ApJ...518..380M}. Each bin has a width of 0.04 mag in $\rm G_{BP}-G_{RP}$, and a height of 0.14 mag in $\rm M_G$. To avoid contamination from white dwarfs on the Q branch and faint white dwarfs, and to only take into account the objects on the bifurcation, we performed the $\chi^2$ statistical test exclusively on the region delimited by the dashed white lines. By electing this region we also avoid uncertainties arising from the star formation rate and the age assumed for the population.

\begin{figure*}
        \centering
        \includegraphics[width=1.\columnwidth,trim={20 5 38 30},clip]{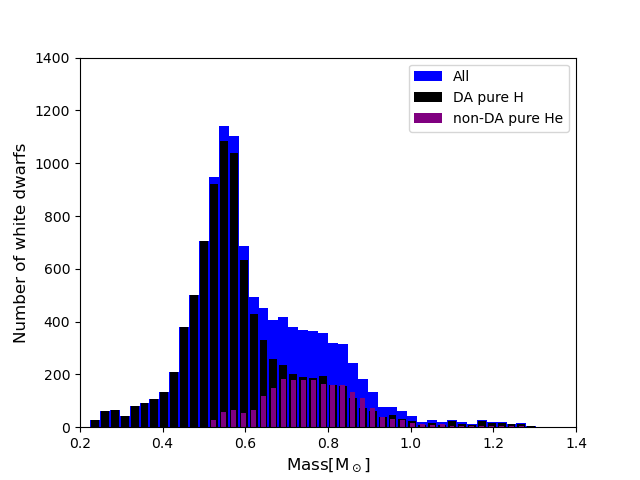}
        \includegraphics[width=1.\columnwidth,trim={20 5 38 30},clip]{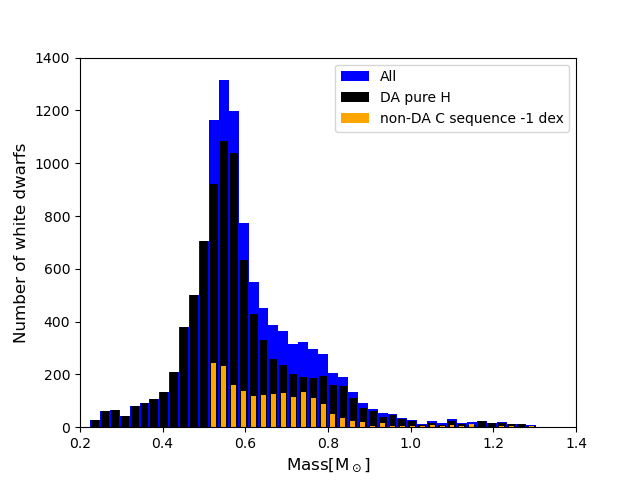}
        \caption{Mass distributions of the 100 pc white dwarf population spectrally classified in DA and non-DA by \cite{2023MNRAS.518.5106J}. Left panel: We employed a pure H atmosphere model for all DA white dwarfs and a pure He atmosphere model for all non-DA white dwarfs. The black, purple and blue histograms are the mass distributions of DA, non-DA and all white dwarfs, respectively.
        Right panel: Same as left panel, but considering a C enrichment following the C sequence -1 dex for all non-DA white dwarfs. The orange histogram shows the mass distribution of these non-DA white dwarfs.} 
        \label{massdistributions}
\end{figure*}

We compared the observed sample with a total of 22 synthetic white dwarf populations varying the proportion of DA to non-DA white dwarfs, the C enrichment prescription for non-DA white dwarfs, and the IFMR. The quantities assumed for our synthetic populations are summarized in Table~\ref{table1}, together with the $\chi^2$ value of the comparison with the observed sample.

 In general lines, we find a better agreement when we include a proportion of 80\% DA, 20\% non-DA, although populations with 75\% DA, 25\% and 70\% DA, 30\% non-DA white dwarfs also show a good agreement. Additionally, the IFMR of \cite{CatalanIFMR} yields a better reproduction of the bifurcation than the one by \cite{BadryIFMR}, even when only DA white dwarfs are considered. A simple synthetic population considering 80\% DA, 20\% non-DA, pure He composition for all non-DA white dwarfs and an IFMR of \cite{CatalanIFMR} is shown in Fig.~\ref{8020pureHe}. We note that this synthetic population does not reproduce the {\it Gaia} bifurcation and has a relatively high value of $\chi^2=3.816$. A similar synthetic population model, with the only difference that it considers an IFMR of \cite{BadryIFMR}, is shown in Fig.~\ref{8020pureHebadry}. Considering this IFMR produces much more white dwarfs with masses $\rm \sim 0.8$\,M$_\odot$, but does not reproduce the {\it Gaia} bifurcation, being its $\chi^2$ value as high as 7.863. 
 On the contrary, considering the PG1159-DO-DB-DQ spectral evolution in all non-DA yields a much better agreement with the observations. In particular, we found the maximum agreement, i.e., the lowest $\chi^2$ value, when we consider a C enrichment that follows the C sequence ($\chi^2=2.972$). In Fig.~\ref{8020Crandom2} we show a synthetic population model that considers 80\% DA, 20\% non-DA, an IFMR of \cite{CatalanIFMR}, pure He composition for all non-DA white dwarfs with $T_\mathrm{eff}>12\,000 \, \mathrm{K}$ and a random C enrichment prescription in the range [0:2] for all non-DA white dwarfs with $T_\mathrm{eff}<12\,000 \, \mathrm{K}$. We note in this figure a slight bifurcation, caused by the C enrichment, that resembles the {\it Gaia} bifurcation. The comparison of this synthetic population model with the observations has $\chi^2=3.042$, which is slightly higher than the value for the best-fit population ($\chi^2=2.972$). Nevertheless, this synthetic population model is more realistic as it takes into account the fact that not all He-dominated white dwarfs can follow the C sequence. If all white dwarfs with He-dominated atmospheres would follow the C sequence, there would be an enormous number of DQ white dwarfs observed, which is not the case. Synthetic population models
 that consist of 80\% DA and 20\% non-DA white dwarfs, following an IFMR of \cite{CatalanIFMR}, and including C enrichment as the C sequence -1 dex or as C random [0:3] also provide a good fit, with $\chi^2$ values of 3.191 and 3.147, respectively.
Therefore, we conclude that, in general  terms, we find much better agreement with the observations for synthetic population models  that consider that non-DA white dwarfs follow a C enrichment, than for synthetic population models that consider that all non-DA white dwarfs have a pure He envelope.

\subsection{The mass distribution}

On the basis of the spectral classification of
\cite{2023MNRAS.518.5106J}, we determined the masses of $11\,455$ DA and $2295$ non-DA white dwarfs within 100\,pc. For each white dwarf, having 
$\rm G$ and $\rm G_{BP}- G_{RP}$, we made a linear interpolation in our evolutionary models to obtain its mass and effective temperature. In order to avoid extrapolation uncertainties, we excluded all DA WDs
with masses lower than $0.239$\,M$_\odot$ and all non-DA WDs with masses lower than $0.51$\,M$_\odot$. For DA white dwarfs, we employed pure H atmosphere models, whereas for non-DA white dwarfs we  determined two different masses, varying the atmosphere model. In the first estimation, we consider that all non-DA white dwarfs have a pure He atmosphere and in the second one, we considered a pure He atmosphere if $T_\mathrm{eff}>12\,000 \mathrm{K}$  and a C enrichment following the C sequence -1 dex if $ T_\mathrm{eff}>12\,000 \mathrm{K}$.

The mass distributions obtained are shown in Fig.~\ref{massdistributions}. The black histograms in both panels show the mass distributions of DA white dwarfs. 
It exhibits a clear peak at $\sim 0.57$\,M$_\odot$, consistent with previous studies. We find a 
slight excess of DA
white dwarfs with masses
$\sim 0.8$\,M$_\odot$ as suggested by \cite{BadryIFMR,2018MNRAS.479L.113K}, but we do also find a flattening in the DA mass distribution around $0.75$\,M$_\odot$. Finally, we do not find a distinctive high-mass excess near but rather some inkling near $1.1$\,M$_\odot$ as found in \cite{2015MNRAS.452.1637R, 
2018MNRAS.480.3942H}.

The mass distributions of non-DA white dwarfs are shown as purple and orange histograms on the left and right panels of Fig.~\ref{massdistributions}, respectively. In the left panel, we employed pure He atmosphere models for all non-DA white dwarfs, whereas in the right panel, we considered the C enrichment commencing at $T_\mathrm{eff}=12\,000$\,K and following the C sequence -1 dex for all non-DA white dwarfs (i.e. ``stealth DQ'' white dwarfs are considered).
The differences arising by the adoption of these treatments can be seen when comparing the purple and orange histograms in this figure. While considering a stealth C enrichment in the atmosphere (orange histogram) leads to a mass distribution more similar to the one for DA white dwarfs, considering a pure He atmosphere (purple histogram) leads to an excess of massive non-DA white dwarfs. Indeed, a surprising massive population of non-DA white dwarfs is obtained when pure He atmosphere models are considered. Such a massive population is not consistent with an isolated evolutionary channel for the formation of non-DA white dwarfs. 
The total mass distributions (DA + non-DA histograms) are shown as blue histograms in both panels. It is clear that considering a pure He envelope leads to an excess of massive ($\sim 0.75$\,M$_\odot$) white dwarfs caused by non-DA white dwarfs.

\section{Summary and conclusions}
\label{conclusions}

The precise observations by {\it Gaia} space mission have revealed a bifurcation in two branches on the color magnitude diagram of the white dwarf population. The main goal of this paper is to provide an explanation for the B branch by investigating the effect of C contamination in the envelopes of He-rich white dwarfs.

There is significant evidence that C contamination occurs on cool He-rich white dwarfs as a result of convective dredge-up, in the so-called PG1159-DO-DB-DQ spectral evolutionary channel. Theoretical models of He-dominated white dwarfs predict that an outer convective zone penetrates into C rich layers, thus leading to C-dredge up and the consequent grow of the surface C abundance. After the convective zone reaches its maximum depth, the partial recombination of C 
below the convective zone makes C sink back into the interior, leading to a slow and constant decrease in the surface C abundance. Although the models can predict the general behaviour of C enrichment, they cannot predict the exact amount of C dredged-up, as it depends on the initial conditions and physical inputs considered. 
Relying on precise He-rich evolutionary models, we simulated the C enrichment under different prescriptions. We considered that all non-DA white dwarfs have a pure He atmosphere if their effective temperature is $>12\,000$K, and a He atmosphere with C traces if their effective temperature is $<12\,000$K. The first C contamination recipe consisted in applying a least-squares fit to the C to He abundance ratio in terms of the effective temperature, observed in cold ($T_\mathrm{eff}<10\,000$\, K) DQ white dwarfs. For $12\,000 \,\mathrm{K}>T_\mathrm{eff}>10\,000\,\mathrm{K}$, we reflected this linear fit (see Sect. \ref{carbonenrichment} for details). We call this type of C contamination ``the C sequence''. Three other similar prescriptions for C contamination have also been considered, by shifting the surface [C/He] abundance from the C sequence by -1 dex, -2 dex, and -3 dex.  It is important to remark that white dwarfs that follow the C sequence enrichment have detectable traces of C in their optical spectra, but white dwarfs that follow the C sequence enrichment -1 dex, -2 dex, and -3 dex do not. Therefore, we call these latter stars ``stealth DQ'' white dwarfs.

We found that the presence of trace C in the atmosphere of He-rich white dwarfs has an important effect on the continuum spectrum of these stars, enhancing the absorption in red wavelengths and thus creating the {\it Gaia} bifurcation. Indeed, the B branch of the {\it Gaia} bifurcation is consistent with a $\sim 0.6$\,M$_\odot$ ``stealth DQ'' evolutionary track. This shift is primarily caused by the He$^-$ free-free absorption. The partial ionisation of trace C leads to a substantial increase in the free electron density, leading to a higher number of He$^-$ ions, amplifying the He$^-$ free-free opacity. Therefore, even though ``stealth DQ'' white dwarfs do not show C lines nor C molecular bands in their optical spectra, their continuum emission differs from the one expected for a pure He atmosphere.
However, ``stealth DQ'' white dwarfs should have strong C signatures in their UV spectra, which would confirm the presence of trace C in their atmospheres.

We performed a population synthesis analysis of the white dwarfs on the {\it Gaia} bifurcation within 100\,pc. We generated synthetic population models varying the IFMR, the DA to non-DA proportion and the non-DA atmospheric composition. We found the maximum agreement with the observations when non-DA models that take into account the C enrichment are considered in the synthetic population models. Non-DA models with a pure He atmosphere fail to reproduce the B branch, even when the flatter IFMR of \cite{BadryIFMR} is considered.
Among the different C enrichment prescriptions, we do not find a significantly better agreement for any prescription in particular. The best $\chi^2$ value was found when non-DA white dwarfs followed the C sequence parametrization, but the C sequence  - 1 dex parametrization also yields a good agreement. Furthermore, the populations that consider that each white dwarf can randomly follow a different C enrichment parametrization also exhibit a good agreement with the observations. We wish to remark that a synthetic population model in which all non-DA white dwarfs follow the C sequence enrichment is not realistic, as it is characterized by a large number of DQ white dwarfs, not detected in observed samples.

Finally, on the basis of the spectral classification of \cite{2023MNRAS.518.5106J}, we determined the white dwarf mass distributions of DA and non-DA white dwarfs within 100\,pc. Having $\rm G$ and $\rm G_{BP}-G_{RP}$ for each individual white dwarf, we interpolated in our white dwarf evolutionary models to obtain its mass. For DA white dwarfs we obtain a peak at $\sim 0.6$\,M$_\odot$ and a flattening around $\sim 0.8$\,M$_\odot$, in agreement with other mass distributions in the literature. For non-DA white dwarfs we obtain two markedly different mass distributions, depending whether we consider a pure He atmosphere or a He atmosphere with traces of C. On the one hand, if a pure He atmosphere is employed, the mass distribution exhibits a wide peak at $\sim 0.8$\,M$_\odot$, which is not consistent with the DA mass distribution and with the standard evolutionary channel for the formation of non-DA white dwarfs (i.e. late thermal pulses). On the other hand, when a mixed He/C atmosphere with ``invisible'' C traces  is employed, we re-obtain the typical peak at $\sim 0.6$\,M$_\odot$. 

In general, we find that the ``stealth DQ'' white dwarfs do a much better job at reproducing bifurcation in the {\it Gaia} color magnitude diagram and the mass distribution for non-DA white dwarfs than pure He white dwarfs. Therefore, we propose that many of the spectrally-classified DC white dwarfs in the B branch of the {\it Gaia} bifurcation could be ``stealth DQ'' white dwarfs having invisible trace C in their optical spectra. Furthermore, \cite{2020A&A...635A.103K} estimated a He mass fraction in the envelope of cold DQs that is nearly one order of magnitude lower than the one predicted by stellar evolutionary models. Hence, ``stealth DQ'' white dwarfs should be originated from white dwarfs with canonical He envelopes, which are consistent with the standard progenitor evolution.
The trace C in some of these stars could potentially be detected in UV spectra. Since only a few DC white dwarfs have been followed up with UV spectra, we encourage observational efforts in detecting C features on the UV spectra of white dwarfs on the B branch of the {\it Gaia} bifurcation.

\begin{acknowledgements}

 The authors acknowledge the expert referee S. O. Kepler for his constructive report.
MC acknowledges
grant RYC2021-032721-I, funded by MCIN/AEI/10.13039/501100011033 and by the European Union NextGenerationEU/PRTR.
MAH is supported by grant ST/V000853/1 from the Science and Technology Facilities Council (STFC). RR acknowledges support from Grant RYC2021-030837-I funded by MCIN/AEI/ 10.13039/501100011033 and by “European Union NextGenerationEU/PRTR”. This work was partially supported by the AGAUR/Generalitat de Catalunya grant SGR-386/2021 and by the Spanish MINECO grant PID2020-117252GB-I00.
 This  research  made use of  NASA Astrophysics Data System. This work made use of data from the European Space Agency (ESA) mission {\it Gaia} (\url{https://www.cosmos.esa.int/gaia}), processed by the {\it Gaia} Data Processing and Analysis Consortium (DPAC, \url{https://www.cosmos.esa.int/web/gaia/dpac/consortium}). Funding for the DPAC has been provided by national institutions, in particular the institutions participating in the {\it Gaia} Multilateral Agreement. 
 
\end{acknowledgements}

\bibliographystyle{aa}
\bibliography{ultramassiveCO}

\begin{thebibliography}{71}
\expandafter\ifx\csname natexlab\endcsname\relax\def\natexlab#1{#1}\fi

\bibitem[{{Althaus} {et~al.}(2015){Althaus}, {Camisassa}, {Miller Bertolami},
  {C{\'o}rsico}, \& {Garc{\'{\i}}a-Berro}}]{2015A&A...576A...9A}
{Althaus}, L.~G., {Camisassa}, M.~E., {Miller Bertolami}, M.~M., {C{\'o}rsico},
  A.~H., \& {Garc{\'{\i}}a-Berro}, E. 2015, \aap, 576, A9

\bibitem[{{Althaus} {et~al.}(2010){Althaus}, {C{\'o}rsico}, {Isern}, \&
  {Garc{\'{\i}}a-Berro}}]{2010A&ARv..18..471A}
{Althaus}, L.~G., {C{\'o}rsico}, A.~H., {Isern}, J., \& {Garc{\'{\i}}a-Berro},
  E. 2010, \aapr, 18, 471

\bibitem[{{Althaus} {et~al.}(2009){Althaus}, {Panei}, {Miller Bertolami},
  {Garc{\'\i}a-Berro}, {C{\'o}rsico}, {Romero}, {Kepler}, \&
  {Rohrmann}}]{2009ApJ...704.1605A}
{Althaus}, L.~G., {Panei}, J.~A., {Miller Bertolami}, M.~M., {et~al.} 2009,
  \apj, 704, 1605

\bibitem[{{Althaus} {et~al.}(2005){Althaus}, {Serenelli}, {Panei},
  {C{\'o}rsico}, {Garc{\'{\i}}a-Berro}, \&
  {Sc{\'o}ccola}}]{2005A&A...435..631A}
{Althaus}, L.~G., {Serenelli}, A.~M., {Panei}, J.~A., {et~al.} 2005, \aap, 435,
  631

\bibitem[{{Bauer} {et~al.}(2020){Bauer}, {Schwab}, {Bildsten}, \&
  {Cheng}}]{2020ApJ...902...93B}
{Bauer}, E.~B., {Schwab}, J., {Bildsten}, L., \& {Cheng}, S. 2020, \apj, 902,
  93

\bibitem[{{B{\'e}dard} {et~al.}(2022){B{\'e}dard}, {Bergeron}, \&
  {Brassard}}]{BedardDQ}
{B{\'e}dard}, A., {Bergeron}, P., \& {Brassard}, P. 2022, \apj, 930, 8

\bibitem[{{B{\'e}dard} {et~al.}(2023){B{\'e}dard}, {Bergeron}, \&
  {Brassard}}]{2023arXiv230205424B}
---. 2023, arXiv e-prints, arXiv:2302.05424

\bibitem[{{B{\'e}dard} {et~al.}(2020){B{\'e}dard}, {Bergeron}, {Brassard}, \&
  {Fontaine}}]{2020ApJ...901...93B}
{B{\'e}dard}, A., {Bergeron}, P., {Brassard}, P., \& {Fontaine}, G. 2020, \apj,
  901, 93

\bibitem[{{Bergeron} {et~al.}(2019){Bergeron}, {Dufour}, {Fontaine}, {Coutu},
  {Blouin}, {Genest-Beaulieu}, {B{\'e}dard}, \& {Rolland}}]{Bergeron2019}
{Bergeron}, P., {Dufour}, P., {Fontaine}, G., {et~al.} 2019, \apj, 876, 67

\bibitem[{{Blouin} {et~al.}(2021){Blouin}, {Daligault}, \&
  {Saumon}}]{2021ApJ...911L...5B}
{Blouin}, S., {Daligault}, J., \& {Saumon}, D. 2021, \apjl, 911, L5

\bibitem[{{Blouin} {et~al.}(2019){Blouin}, {Dufour}, {Thibeault}, \&
  {Allard}}]{2019ApJ...878...63B}
{Blouin}, S., {Dufour}, P., {Thibeault}, C., \& {Allard}, N.~F. 2019, \apj,
  878, 63

\bibitem[{{Camacho} {et~al.}(2014){Camacho}, {Torres}, {Garc{\'\i}a-Berro},
  {Zorotovic}, {Schreiber}, {Rebassa-Mansergas}, {Nebot G{\'o}mez-Mor{\'a}n},
  \& {G{\"a}nsicke}}]{Camacho2014}
{Camacho}, J., {Torres}, S., {Garc{\'\i}a-Berro}, E., {et~al.} 2014, \aap, 566,
  A86

\bibitem[{{Camisassa} {et~al.}(2019){Camisassa}, {Althaus}, {C{\'o}rsico}, {De
  Ger{\'o}nimo}, {Miller Bertolami}, {Novarino}, {Rohrmann}, {Wachlin}, \&
  {Garc{\'\i}a-Berro}}]{2019A&A...625A..87C}
{Camisassa}, M.~E., {Althaus}, L.~G., {C{\'o}rsico}, A.~H., {et~al.} 2019,
  \aap, 625, A87

\bibitem[{{Camisassa} {et~al.}(2016){Camisassa}, {Althaus}, {C{\'o}rsico},
  {Vinyoles}, {Serenelli}, {Isern}, {Miller Bertolami}, \&
  {Garc{\'\i}a─Berro}}]{Camisassa2016}
---. 2016, \apj, 823, 158

\bibitem[{{Camisassa} {et~al.}(2022){Camisassa}, {Althaus}, {Koester},
  {Torres}, {Pons}, \& {C{\'o}rsico}}]{2022MNRAS.511.5198C}
{Camisassa}, M.~E., {Althaus}, L.~G., {Koester}, D., {et~al.} 2022, \mnras,
  511, 5198

\bibitem[{{Camisassa} {et~al.}(2017){Camisassa}, {Althaus}, {Rohrmann},
  {Garc{\'\i}a-Berro}, {Torres}, {C{\'o}rsico}, \& {Wachlin}}]{Camisassa2017}
{Camisassa}, M.~E., {Althaus}, L.~G., {Rohrmann}, R.~D., {et~al.} 2017, \apj,
  839, 11

\bibitem[{{Camisassa} {et~al.}(2021){Camisassa}, {Althaus}, {Torres},
  {C{\'o}rsico}, {Rebassa-Mansergas}, {Tremblay}, {Cheng}, \&
  {Raddi}}]{2021A&A...649L...7C}
{Camisassa}, M.~E., {Althaus}, L.~G., {Torres}, S., {et~al.} 2021, \aap, 649,
  L7

\bibitem[{{Canals} {et~al.}(2018){Canals}, {Torres}, \&
  {Soker}}]{2018MNRAS.480.4519C}
{Canals}, P., {Torres}, S., \& {Soker}, N. 2018, \mnras, 480, 4519

\bibitem[{{Caplan} {et~al.}(2021){Caplan}, {Freeman}, {Horowitz}, {Cumming}, \&
  {Bellinger}}]{2021ApJ...919L..12C}
{Caplan}, M.~E., {Freeman}, I.~F., {Horowitz}, C.~J., {Cumming}, A., \&
  {Bellinger}, E.~P. 2021, \apjl, 919, L12

\bibitem[{{Catal{\'a}n} {et~al.}(2008){Catal{\'a}n}, {Isern},
  {Garc{\'\i}a-Berro}, \& {Ribas}}]{CatalanIFMR}
{Catal{\'a}n}, S., {Isern}, J., {Garc{\'\i}a-Berro}, E., \& {Ribas}, I. 2008,
  \mnras, 387, 1693

\bibitem[{{Chambers} {et~al.}(2016){Chambers}, {Magnier}, {Metcalfe},
  {Flewelling}, {Huber}, {Waters}, {Denneau}, {Draper}, {Farrow}, {Finkbeiner},
  {Holmberg}, {Koppenhoefer}, {Price}, {Rest}, {Saglia}, {Schlafly}, {Smartt},
  {Sweeney}, {Wainscoat}, {Burgett}, {Chastel}, {Grav}, {Heasley}, {Hodapp},
  {Jedicke}, {Kaiser}, {Kudritzki}, {Luppino}, {Lupton}, {Monet}, {Morgan},
  {Onaka}, {Shiao}, {Stubbs}, {Tonry}, {White}, {Ba{\~n}ados}, {Bell},
  {Bender}, {Bernard}, {Boegner}, {Boffi}, {Botticella}, {Calamida},
  {Casertano}, {Chen}, {Chen}, {Cole}, {Deacon}, {Frenk}, {Fitzsimmons},
  {Gezari}, {Gibbs}, {Goessl}, {Goggia}, {Gourgue}, {Goldman}, {Grant},
  {Grebel}, {Hambly}, {Hasinger}, {Heavens}, {Heckman}, {Henderson}, {Henning},
  {Holman}, {Hopp}, {Ip}, {Isani}, {Jackson}, {Keyes}, {Koekemoer}, {Kotak},
  {Le}, {Liska}, {Long}, {Lucey}, {Liu}, {Martin}, {Masci}, {McLean}, {Mindel},
  {Misra}, {Morganson}, {Murphy}, {Obaika}, {Narayan}, {Nieto-Santisteban},
  {Norberg}, {Peacock}, {Pier}, {Postman}, {Primak}, {Rae}, {Rai}, {Riess},
  {Riffeser}, {Rix}, {R{\"o}ser}, {Russel}, {Rutz}, {Schilbach}, {Schultz},
  {Scolnic}, {Strolger}, {Szalay}, {Seitz}, {Small}, {Smith}, {Soderblom},
  {Taylor}, {Thomson}, {Taylor}, {Thakar}, {Thiel}, {Thilker}, {Unger},
  {Urata}, {Valenti}, {Wagner}, {Walder}, {Walter}, {Watters}, {Werner},
  {Wood-Vasey}, \& {Wyse}}]{2016arXiv161205560C}
{Chambers}, K.~C., {Magnier}, E.~A., {Metcalfe}, N., {et~al.} 2016, arXiv
  e-prints, arXiv:1612.05560

\bibitem[{{Cheng} {et~al.}(2019){Cheng}, {Cummings}, \&
  {M{\'e}nard}}]{Cheng2019}
{Cheng}, S., {Cummings}, J.~D., \& {M{\'e}nard}, B. 2019, \apj, 886, 100

\bibitem[{{Cojocaru} {et~al.}(2017){Cojocaru}, {Rebassa-Mansergas}, {Torres},
  \& {Garc{\'\i}a-Berro}}]{Cojocaru2017}
{Cojocaru}, R., {Rebassa-Mansergas}, A., {Torres}, S., \& {Garc{\'\i}a-Berro},
  E. 2017, \mnras, 470, 1442

\bibitem[{{C{\'o}rsico} {et~al.}(2019){C{\'o}rsico}, {Althaus}, {Miller
  Bertolami}, \& {Kepler}}]{2019A&ARv..27....7C}
{C{\'o}rsico}, A.~H., {Althaus}, L.~G., {Miller Bertolami}, M.~M., \& {Kepler},
  S.~O. 2019, \aapr, 27, 7

\bibitem[{{Dufour} {et~al.}(2005){Dufour}, {Bergeron}, \&
  {Fontaine}}]{2005ApJ...627..404D}
{Dufour}, P., {Bergeron}, P., \& {Fontaine}, G. 2005, \apj, 627, 404

\bibitem[{{Dunlap} \& {Clemens}(2015)}]{2015ASPC..493..547D}
{Dunlap}, B.~H. \& {Clemens}, J.~C. 2015, in Astronomical Society of the
  Pacific Conference Series, Vol. 493, 19th European Workshop on White Dwarfs,
  ed. P.~{Dufour}, P.~{Bergeron}, \& G.~{Fontaine}, 547

\bibitem[{{El-Badry} {et~al.}(2018){El-Badry}, {Rix}, \& {Weisz}}]{BadryIFMR}
{El-Badry}, K., {Rix}, H.-W., \& {Weisz}, D.~R. 2018, \apjl, 860, L17

\bibitem[{{Fleury} {et~al.}(2022){Fleury}, {Caiazzo}, \&
  {Heyl}}]{2022MNRAS.511.5984F}
{Fleury}, L., {Caiazzo}, I., \& {Heyl}, J. 2022, \mnras, 511, 5984

\bibitem[{{Fontaine} \& {Brassard}(2008)}]{2008PASP..120.1043F}
{Fontaine}, G. \& {Brassard}, P. 2008, PASP, 120, 1043

\bibitem[{{Gaia Collaboration} {et~al.}(2018){Gaia Collaboration}, {Babusiaux},
  {van Leeuwen}, {Barstow}, {Jordi}, {Vallenari}, {Bossini}, {Bressan},
  {Cantat-Gaudin}, {van Leeuwen}, {Brown}, {Prusti}, {de Bruijne},
  {Bailer-Jones}, {Biermann}, {Evans}, {Eyer}, {Jansen}, {Klioner}, {Lammers},
  {Lindegren}, {Luri}, {Mignard}, {Panem}, {Pourbaix}, {Randich}, {Sartoretti},
  {Siddiqui}, {Soubiran}, {Walton}, {Arenou}, {Bastian}, {Cropper}, {Drimmel},
  {Katz}, {Lattanzi}, {Bakker}, {Cacciari}, {Casta{\~n}eda}, {Chaoul}, {Cheek},
  {De Angeli}, {Fabricius}, {Guerra}, {Holl}, {Masana}, {Messineo}, {Mowlavi},
  {Nienartowicz}, {Panuzzo}, {Portell}, {Riello}, {Seabroke}, {Tanga},
  {Th{\'e}venin}, {Gracia-Abril}, {Comoretto}, {Garcia-Reinaldos}, {Teyssier},
  {Altmann}, {Andrae}, {Audard}, {Bellas-Velidis}, {Benson}, {Berthier},
  {Blomme}, {Burgess}, {Busso}, {Carry}, {Cellino}, {Clementini}, {Clotet},
  {Creevey}, {Davidson}, {De Ridder}, {Delchambre}, {Dell'Oro}, {Ducourant},
  {Fern{\'a}ndez-Hern{\'a}ndez}, {Fouesneau}, {Fr{\'e}mat}, {Galluccio},
  {Garc{\'\i}a-Torres}, {Gonz{\'a}lez-N{\'u}{\~n}ez}, {Gonz{\'a}lez-Vidal},
  {Gosset}, {Guy}, {Halbwachs}, {Hambly}, {Harrison}, {Hern{\'a}ndez},
  {Hestroffer}, {Hodgkin}, {Hutton}, {Jasniewicz}, {Jean-Antoine-Piccolo},
  {Jordan}, {Korn}, {Krone-Martins}, {Lanzafame}, {Lebzelter}, {L{\"o}ffler},
  {Manteiga}, {Marrese}, {Mart{\'\i}n-Fleitas}, {Moitinho}, {Mora}, {Muinonen},
  {Osinde}, {Pancino}, {Pauwels}, {Petit}, {Recio-Blanco}, {Richards},
  {Rimoldini}, {Robin}, {Sarro}, {Siopis}, {Smith}, {Sozzetti}, {S{\"u}veges},
  {Torra}, {van Reeven}, {Abbas}, {Abreu Aramburu}, {Accart}, {Aerts},
  {Altavilla}, {{\'A}lvarez}, {Alvarez}, {Alves}, {Anderson}, {Andrei},
  {Anglada Varela}, {Antiche}, {Antoja}, {Arcay}, {Astraatmadja}, {Bach},
  {Baker}, {Balaguer-N{\'u}{\~n}ez}, {Balm}, {Barache}, {Barata}, {Barbato},
  {Barblan}, {Barklem}, {Barrado}, {Barros}, {Bartholom{\'e} Mu{\~n}oz},
  {Bassilana}, {Becciani}, {Bellazzini}, {Berihuete}, {Bertone}, {Bianchi},
  {Bienaym{\'e}}, {Blanco-Cuaresma}, {Boch}, {Boeche}, {Bombrun}, {Borrachero},
  {Bouquillon}, {Bourda}, {Bragaglia}, {Bramante}, {Breddels}, {Brouillet},
  {Br{\"u}semeister}, {Brugaletta}, {Bucciarelli}, {Burlacu}, {Busonero},
  {Butkevich}, {Buzzi}, {Caffau}, {Cancelliere}, {Cannizzaro}, {Carballo},
  {Carlucci}, {Carrasco}, {Casamiquela}, {Castellani}, {Castro-Ginard},
  {Charlot}, {Chemin}, {Chiavassa}, {Cocozza}, {Costigan}, {Cowell}, {Crifo},
  {Crosta}, {Crowley}, {Cuypers}, {Dafonte}, {Damerdji}, {Dapergolas}, {David},
  {David}, {de Laverny}, {De Luise}, {De March}, {de Martino}, {de Souza}, {de
  Torres}, {Debosscher}, {del Pozo}, {Delbo}, {Delgado}, {Delgado}, {Diakite},
  {Diener}, {Distefano}, {Dolding}, {Drazinos}, {Dur{\'a}n}, {Edvardsson},
  {Enke}, {Eriksson}, {Esquej}, {Eynard Bontemps}, {Fabre}, {Fabrizio},
  {Faigler}, {Falc{\~a}o}, {Farr{\`a}s Casas}, {Federici}, {Fedorets},
  {Fernique}, {Figueras}, {Filippi}, {Findeisen}, {Fonti}, {Fraile}, {Fraser},
  {Fr{\'e}zouls}, {Gai}, {Galleti}, {Garabato}, {Garc{\'\i}a-Sedano},
  {Garofalo}, {Garralda}, {Gavel}, {Gavras}, {Gerssen}, {Geyer}, {Giacobbe},
  {Gilmore}, {Girona}, {Giuffrida}, {Glass}, {Gomes}, {Granvik}, {Gueguen},
  {Guerrier}, {Guiraud}, {Guti{\'e}}, {Haigron}, {Hatzidimitriou}, {Hauser},
  {Haywood}, {Heiter}, {Helmi}, {Heu}, {Hilger}, {Hobbs}, {Hofmann}, {Holland},
  {Huckle}, {Hypki}, {Icardi}, {Jan{\ss}en}, {Jevardat de Fombelle}, {Jonker},
  {Juh{\'a}sz}, {Julbe}, {Karampelas}, {Kewley}, {Klar}, {Kochoska}, {Kohley},
  {Kolenberg}, {Kontizas}, {Kontizas}, {Koposov}, {Kordopatis},
  {Kostrzewa-Rutkowska}, {Koubsky}, {Lambert}, {Lanza}, {Lasne}, {Lavigne}, {Le
  Fustec}, {Le Poncin-Lafitte}, {Lebreton}, {Leccia}, {Leclerc},
  {Lecoeur-Taibi}, {Lenhardt}, {Leroux}, {Liao}, {Licata}, {Lindstr{\o}m},
  {Lister}, {Livanou}, {Lobel}, {L{\'o}pez}, {Managau}, {Mann}, {Mantelet},
  {Marchal}, {Marchant}, {Marconi}, {Marinoni}, {Marschalk{\'o}}, {Marshall},
  {Martino}, {Marton}, {Mary}, {Massari}, {Matijevi{\v{c}}}, {Mazeh},
  {McMillan}, {Messina}, {Michalik}, {Millar}, {Molina}, {Molinaro},
  {Moln{\'a}r}, {Montegriffo}, {Mor}, {Morbidelli}, {Morel}, {Morris},
  {Mulone}, {Muraveva}, {Musella}, {Nelemans}, {Nicastro}, {Noval},
  {O'Mullane}, {Ord{\'e}novic}, {Ord{\'o}{\~n}ez-Blanco}, {Osborne}, {Pagani},
  {Pagano}, {Pailler}, {Palacin}, {Palaversa}, {Panahi}, {Pawlak},
  {Piersimoni}, {Pineau}, {Plachy}, {Plum}, {Poggio}, {Poujoulet},
  {Pr{\v{s}}a}, {Pulone}, {Racero}, {Ragaini}, {Rambaux}, {Ramos-Lerate},
  {Regibo}, {Reyl{\'e}}, {Riclet}, {Ripepi}, {Riva}, {Rivard}, {Rixon},
  {Roegiers}, {Roelens}, {Romero-G{\'o}mez}, {Rowell}, {Royer}, {Ruiz-Dern},
  {Sadowski}, {Sagrist{\`a} Sell{\'e}s}, {Sahlmann}, {Salgado}, {Salguero},
  {Sanna}, {Santana-Ros}, {Sarasso}, {Savietto}, {Schultheis}, {Sciacca},
  {Segol}, {Segovia}, {S{\'e}gransan}, {Shih}, {Siltala}, {Silva}, {Smart},
  {Smith}, {Solano}, {Solitro}, {Sordo}, {Soria Nieto}, {Souchay}, {Spagna},
  {Spoto}, {Stampa}, {Steele}, {Steidelm{\"u}ller}, {Stephenson}, {Stoev},
  {Suess}, {Surdej}, {Szabados}, {Szegedi-Elek}, {Tapiador}, {Taris}, {Tauran},
  {Taylor}, {Teixeira}, {Terrett}, {Teyssandier}, {Thuillot}, {Titarenko},
  {Torra Clotet}, {Turon}, {Ulla}, {Utrilla}, {Uzzi}, {Vaillant}, {Valentini},
  {Valette}, {van Elteren}, {Van Hemelryck}, {Vaschetto}, {Vecchiato},
  {Veljanoski}, {Viala}, {Vicente}, {Vogt}, {von Essen}, {Voss}, {Votruba},
  {Voutsinas}, {Walmsley}, {Weiler}, {Wertz}, {Wevers}, {Wyrzykowski},
  {Yoldas}, {{\v{Z}}erjal}, {Ziaeepour}, {Zorec}, {Zschocke}, {Zucker},
  {Zurbach}, \& {Zwitter}}]{GaiaDR22018}
{Gaia Collaboration}, {Babusiaux}, C., {van Leeuwen}, F., {et~al.} 2018, \aap,
  616, A10

\bibitem[{{Gaia Collaboration} {et~al.}(2021){Gaia Collaboration}, {Brown},
  {Vallenari}, {Prusti}, {de Bruijne}, {Babusiaux}, {Biermann}, {Creevey},
  {Evans}, {Eyer}, {Hutton}, {Jansen}, {Jordi}, {Klioner}, {Lammers},
  {Lindegren}, {Luri}, {Mignard}, {Panem}, {Pourbaix}, {Randich}, {Sartoretti},
  {Soubiran}, {Walton}, {Arenou}, {Bailer-Jones}, {Bastian}, {Cropper},
  {Drimmel}, {Katz}, {Lattanzi}, {van Leeuwen}, {Bakker}, {Cacciari},
  {Casta{\~n}eda}, {De Angeli}, {Ducourant}, {Fabricius}, {Fouesneau},
  {Fr{\'e}mat}, {Guerra}, {Guerrier}, {Guiraud}, {Jean-Antoine Piccolo},
  {Masana}, {Messineo}, {Mowlavi}, {Nicolas}, {Nienartowicz}, {Pailler},
  {Panuzzo}, {Riclet}, {Roux}, {Seabroke}, {Sordo}, {Tanga}, {Th{\'e}venin},
  {Gracia-Abril}, {Portell}, {Teyssier}, {Altmann}, {Andrae}, {Bellas-Velidis},
  {Benson}, {Berthier}, {Blomme}, {Brugaletta}, {Burgess}, {Busso}, {Carry},
  {Cellino}, {Cheek}, {Clementini}, {Damerdji}, {Davidson}, {Delchambre},
  {Dell'Oro}, {Fern{\'a}ndez-Hern{\'a}ndez}, {Galluccio}, {Garc{\'\i}a-Lario},
  {Garcia-Reinaldos}, {Gonz{\'a}lez-N{\'u}{\~n}ez}, {Gosset}, {Haigron},
  {Halbwachs}, {Hambly}, {Harrison}, {Hatzidimitriou}, {Heiter},
  {Hern{\'a}ndez}, {Hestroffer}, {Hodgkin}, {Holl}, {Jan{\ss}en}, {Jevardat de
  Fombelle}, {Jordan}, {Krone-Martins}, {Lanzafame}, {L{\"o}ffler}, {Lorca},
  {Manteiga}, {Marchal}, {Marrese}, {Moitinho}, {Mora}, {Muinonen}, {Osborne},
  {Pancino}, {Pauwels}, {Petit}, {Recio-Blanco}, {Richards}, {Riello},
  {Rimoldini}, {Robin}, {Roegiers}, {Rybizki}, {Sarro}, {Siopis}, {Smith},
  {Sozzetti}, {Ulla}, {Utrilla}, {van Leeuwen}, {van Reeven}, {Abbas}, {Abreu
  Aramburu}, {Accart}, {Aerts}, {Aguado}, {Ajaj}, {Altavilla}, {{\'A}lvarez},
  {{\'A}lvarez Cid-Fuentes}, {Alves}, {Anderson}, {Anglada Varela}, {Antoja},
  {Audard}, {Baines}, {Baker}, {Balaguer-N{\'u}{\~n}ez}, {Balbinot}, {Balog},
  {Barache}, {Barbato}, {Barros}, {Barstow}, {Bartolom{\'e}}, {Bassilana},
  {Bauchet}, {Baudesson-Stella}, {Becciani}, {Bellazzini}, {Bernet}, {Bertone},
  {Bianchi}, {Blanco-Cuaresma}, {Boch}, {Bombrun}, {Bossini}, {Bouquillon},
  {Bragaglia}, {Bramante}, {Breedt}, {Bressan}, {Brouillet}, {Bucciarelli},
  {Burlacu}, {Busonero}, {Butkevich}, {Buzzi}, {Caffau}, {Cancelliere},
  {C{\'a}novas}, {Cantat-Gaudin}, {Carballo}, {Carlucci}, {Carnerero},
  {Carrasco}, {Casamiquela}, {Castellani}, {Castro-Ginard}, {Castro Sampol},
  {Chaoul}, {Charlot}, {Chemin}, {Chiavassa}, {Cioni}, {Comoretto}, {Cooper},
  {Cornez}, {Cowell}, {Crifo}, {Crosta}, {Crowley}, {Dafonte}, {Dapergolas},
  {David}, {David}, {de Laverny}, {De Luise}, {De March}, {De Ridder}, {de
  Souza}, {de Teodoro}, {de Torres}, {del Peloso}, {del Pozo}, {Delbo},
  {Delgado}, {Delgado}, {Delisle}, {Di Matteo}, {Diakite}, {Diener},
  {Distefano}, {Dolding}, {Eappachen}, {Edvardsson}, {Enke}, {Esquej}, {Fabre},
  {Fabrizio}, {Faigler}, {Fedorets}, {Fernique}, {Fienga}, {Figueras},
  {Fouron}, {Fragkoudi}, {Fraile}, {Franke}, {Gai}, {Garabato},
  {Garcia-Gutierrez}, {Garc{\'\i}a-Torres}, {Garofalo}, {Gavras}, {Gerlach},
  {Geyer}, {Giacobbe}, {Gilmore}, {Girona}, {Giuffrida}, {Gomel}, {Gomez},
  {Gonzalez-Santamaria}, {Gonz{\'a}lez-Vidal}, {Granvik},
  {Guti{\'e}rrez-S{\'a}nchez}, {Guy}, {Hauser}, {Haywood}, {Helmi}, {Hidalgo},
  {Hilger}, {H{\l}adczuk}, {Hobbs}, {Holland}, {Huckle}, {Jasniewicz},
  {Jonker}, {Juaristi Campillo}, {Julbe}, {Karbevska}, {Kervella}, {Khanna},
  {Kochoska}, {Kontizas}, {Kordopatis}, {Korn}, {Kostrzewa-Rutkowska},
  {Kruszy{\'n}ska}, {Lambert}, {Lanza}, {Lasne}, {Le Campion}, {Le Fustec},
  {Lebreton}, {Lebzelter}, {Leccia}, {Leclerc}, {Lecoeur-Taibi}, {Liao},
  {Licata}, {Lindstr{\o}m}, {Lister}, {Livanou}, {Lobel}, {Madrero Pardo},
  {Managau}, {Mann}, {Marchant}, {Marconi}, {Marcos Santos}, {Marinoni},
  {Marocco}, {Marshall}, {Martin Polo}, {Mart{\'\i}n-Fleitas}, {Masip},
  {Massari}, {Mastrobuono-Battisti}, {Mazeh}, {McMillan}, {Messina},
  {Michalik}, {Millar}, {Mints}, {Molina}, {Molinaro}, {Moln{\'a}r},
  {Montegriffo}, {Mor}, {Morbidelli}, {Morel}, {Morris}, {Mulone}, {Munoz},
  {Muraveva}, {Murphy}, {Musella}, {Noval}, {Ord{\'e}novic}, {Orr{\`u}},
  {Osinde}, {Pagani}, {Pagano}, {Palaversa}, {Palicio}, {Panahi}, {Pawlak},
  {Pe{\~n}alosa Esteller}, {Penttil{\"a}}, {Piersimoni}, {Pineau}, {Plachy},
  {Plum}, {Poggio}, {Poretti}, {Poujoulet}, {Pr{\v{s}}a}, {Pulone}, {Racero},
  {Ragaini}, {Rainer}, {Raiteri}, {Rambaux}, {Ramos}, {Ramos-Lerate}, {Re
  Fiorentin}, {Regibo}, {Reyl{\'e}}, {Ripepi}, {Riva}, {Rixon}, {Robichon},
  {Robin}, {Roelens}, {Rohrbasser}, {Romero-G{\'o}mez}, {Rowell}, {Royer},
  {Rybicki}, {Sadowski}, {Sagrist{\`a} Sell{\'e}s}, {Sahlmann}, {Salgado},
  {Salguero}, {Samaras}, {Sanchez Gimenez}, {Sanna}, {Santove{\~n}a},
  {Sarasso}, {Schultheis}, {Sciacca}, {Segol}, {Segovia}, {S{\'e}gransan},
  {Semeux}, {Shahaf}, {Siddiqui}, {Siebert}, {Siltala}, {Slezak}, {Smart},
  {Solano}, {Solitro}, {Souami}, {Souchay}, {Spagna}, {Spoto}, {Steele},
  {Steidelm{\"u}ller}, {Stephenson}, {S{\"u}veges}, {Szabados}, {Szegedi-Elek},
  {Taris}, {Tauran}, {Taylor}, {Teixeira}, {Thuillot}, {Tonello}, {Torra},
  {Torra}, {Turon}, {Unger}, {Vaillant}, {van Dillen}, {Vanel}, {Vecchiato},
  {Viala}, {Vicente}, {Voutsinas}, {Weiler}, {Wevers}, {Wyrzykowski}, {Yoldas},
  {Yvard}, {Zhao}, {Zorec}, {Zucker}, {Zurbach}, \& {Zwitter}}]{GaiaEDR32021}
{Gaia Collaboration}, {Brown}, A.~G.~A., {Vallenari}, A., {et~al.} 2021, \aap,
  649, A1

\bibitem[{{Garc\'{i}a-Berro} {et~al.}(1999){Garc\'{i}a-Berro}, E., {Isern}, \&
  {Burkert}}]{GBerro1999}
{Garc\'{i}a-Berro}, E., {Torres}, S., {Isern}, J., \& {Burkert}, A. 1999,
  \mnras, 302, 173

\bibitem[{{Garc{\'{\i}}a-Berro} \& {Oswalt}(2016)}]{2016NewAR..72....1G}
{Garc{\'{\i}}a-Berro}, E. \& {Oswalt}, T.~D. 2016, New Astronomy Reviews, 72, 1

\bibitem[{{Garc{\'{\i}}a-Berro} {et~al.}(2010){Garc{\'{\i}}a-Berro}, {Torres},
  {Althaus}, {Renedo}, {Lor{\'e}n-Aguilar}, {C{\'o}rsico}, {Rohrmann},
  {Salaris}, \& {Isern}}]{2010Natur.465..194G}
{Garc{\'{\i}}a-Berro}, E., {Torres}, S., {Althaus}, L.~G., {et~al.} 2010,
  nature, 465, 194

\bibitem[{{Gentile Fusillo} {et~al.}(2021){Gentile Fusillo}, {Tremblay},
  {Cukanovaite}, {Vorontseva}, {Lallement}, {Hollands}, {G{\"a}nsicke},
  {Burdge}, {McCleery}, \& {Jordan}}]{Fusillo2021}
{Gentile Fusillo}, N.~P., {Tremblay}, P.~E., {Cukanovaite}, E., {et~al.} 2021,
  \mnras, 508, 3877

\bibitem[{Hidalgo {et~al.}(2018)Hidalgo, Pietrinferni, Cassisi, Salaris,
  Mucciarelli, Savino, Aparicio, Silva~Aguirre, \& Verma}]{bastinew}
Hidalgo, S., Pietrinferni, A., Cassisi, S., {et~al.} 2018, The Astrophysical
  Journal, 856

\bibitem[{{Hollands} {et~al.}(2018){Hollands}, {Tremblay}, {G{\"a}nsicke},
  {Gentile-Fusillo}, \& {Toonen}}]{2018MNRAS.480.3942H}
{Hollands}, M.~A., {Tremblay}, P.~E., {G{\"a}nsicke}, B.~T., {Gentile-Fusillo},
  N.~P., \& {Toonen}, S. 2018, \mnras, 480, 3942

\bibitem[{{Isern} {et~al.}(2022){Isern}, {Torres}, \&
  {Rebassa-Mansergas}}]{2022FrASS...9....6I}
{Isern}, J., {Torres}, S., \& {Rebassa-Mansergas}, A. 2022, Frontiers in
  Astronomy and Space Sciences, 9, 6

\bibitem[{{Jim{\'e}nez-Esteban} {et~al.}(2023){Jim{\'e}nez-Esteban}, {Torres},
  {Rebassa-Mansergas}, {Cruz}, {Murillo-Ojeda}, {Solano}, {Rodrigo}, \&
  {Camisassa}}]{2023MNRAS.518.5106J}
{Jim{\'e}nez-Esteban}, F.~M., {Torres}, S., {Rebassa-Mansergas}, A., {et~al.}
  2023, \mnras, 518, 5106

\bibitem[{{Jim{\'e}nez-Esteban} {et~al.}(2018){Jim{\'e}nez-Esteban}, {Torres},
  {Rebassa-Mansergas}, {Skorobogatov}, {Solano}, {Cantero}, \&
  {Rodrigo}}]{Jimenez2018}
---. 2018, \mnras, 480, 4505

\bibitem[{{Kawka} {et~al.}(2023){Kawka}, {Ferrario}, \&
  {Vennes}}]{2023MNRAS.520.6299K}
{Kawka}, A., {Ferrario}, L., \& {Vennes}, S. 2023, \mnras, 520, 6299

\bibitem[{{Kepler} {et~al.}(2021){Kepler}, {Koester}, {Pelisoli}, {Romero}, \&
  {Ourique}}]{Kepler2021}
{Kepler}, S.~O., {Koester}, D., {Pelisoli}, I., {Romero}, A.~D., \& {Ourique},
  G. 2021, \mnras, 507, 4646

\bibitem[{{Kilic} {et~al.}(2018){Kilic}, {Hambly}, {Bergeron},
  {Genest-Beaulieu}, \& {Rowell}}]{2018MNRAS.479L.113K}
{Kilic}, M., {Hambly}, N.~C., {Bergeron}, P., {Genest-Beaulieu}, C., \&
  {Rowell}, N. 2018, \mnras, 479, L113

\bibitem[{{Koester}(2010)}]{2010MmSAI..81..921K}
{Koester}, D. 2010, \memsai, 81, 921

\bibitem[{{Koester} \& {Kepler}(2019)}]{2019A&A...628A.102K}
{Koester}, D. \& {Kepler}, S.~O. 2019, \aap, 628, A102

\bibitem[{{Koester} {et~al.}(2020){Koester}, {Kepler}, \&
  {Irwin}}]{2020A&A...635A.103K}
{Koester}, D., {Kepler}, S.~O., \& {Irwin}, A.~W. 2020, \aap, 635, A103

\bibitem[{{Koester} {et~al.}(1982){Koester}, {Weidemann}, \&
  {Zeidler}}]{1982A&A...116..147K}
{Koester}, D., {Weidemann}, V., \& {Zeidler}, E.~M. 1982, \aap, 116, 147

\bibitem[{{Mighell}(1999)}]{1999ApJ...518..380M}
{Mighell}, K.~J. 1999, \apj, 518, 380

\bibitem[{{Miller Bertolami}(2016)}]{2016A&A...588A..25M}
{Miller Bertolami}, M.~M. 2016, \aap, 588, A25

\bibitem[{{Miller Bertolami} \& {Althaus}(2006)}]{2006A&A...454..845M}
{Miller Bertolami}, M.~M. \& {Althaus}, L.~G. 2006, \aap, 454, 845

\bibitem[{{Ourique} {et~al.}(2020){Ourique}, {Kepler}, {Romero}, {Klippel}, \&
  {Koester}}]{2020MNRAS.492.5003O}
{Ourique}, G., {Kepler}, S.~O., {Romero}, A.~D., {Klippel}, T.~S., \&
  {Koester}, D. 2020, \mnras, 492, 5003

\bibitem[{{Pelletier} {et~al.}(1986){Pelletier}, {Fontaine}, {Wesemael},
  {Michaud}, \& {Wegner}}]{1986ApJ...307..242P}
{Pelletier}, C., {Fontaine}, G., {Wesemael}, F., {Michaud}, G., \& {Wegner}, G.
  1986, \apj, 307, 242

\bibitem[{{Raddi} {et~al.}(2022){Raddi}, {Torres}, {Rebassa-Mansergas},
  {Maldonado}, {Camisassa}, {Koester}, {Gentile Fusillo}, {Tremblay}, {Dimpel},
  {Heber}, {Cunningham}, \& {Ren}}]{2022A&A...658A..22R}
{Raddi}, R., {Torres}, S., {Rebassa-Mansergas}, A., {et~al.} 2022, \aap, 658,
  A22

\bibitem[{{Rebassa-Mansergas} {et~al.}(2021){Rebassa-Mansergas}, {Maldonado},
  {Raddi}, {Knowles}, {Torres}, {Hoskin}, {Cunningham}, {Hollands}, {Ren},
  {G{\"a}nsicke}, {Tremblay}, {Castro-Rodr{\'\i}guez}, {Camisassa}, \&
  {Koester}}]{2021MNRAS.505.3165R}
{Rebassa-Mansergas}, A., {Maldonado}, J., {Raddi}, R., {et~al.} 2021, \mnras,
  505, 3165

\bibitem[{{Rebassa-Mansergas} {et~al.}(2015){Rebassa-Mansergas}, {Rybicka},
  {Liu}, {Han}, \& {Garc{\'\i}a-Berro}}]{2015MNRAS.452.1637R}
{Rebassa-Mansergas}, A., {Rybicka}, M., {Liu}, X.~W., {Han}, Z., \&
  {Garc{\'\i}a-Berro}, E. 2015, \mnras, 452, 1637

\bibitem[{{Renedo} {et~al.}(2010){Renedo}, {Althaus}, {Miller Bertolami},
  {Romero}, {C{\'o}rsico}, {Rohrmann}, \&
  {Garc{\'{\i}}a-Berro}}]{2010ApJ...717..183R}
{Renedo}, I., {Althaus}, L.~G., {Miller Bertolami}, M.~M., {et~al.} 2010, \apj,
  717, 183

\bibitem[{{Ricker} {et~al.}(2015){Ricker}, {Winn}, {Vanderspek}, {Latham},
  {Bakos}, {Bean}, {Berta-Thompson}, {Brown}, {Buchhave}, {Butler}, {Butler},
  {Chaplin}, {Charbonneau}, {Christensen-Dalsgaard}, {Clampin}, {Deming},
  {Doty}, {De Lee}, {Dressing}, {Dunham}, {Endl}, {Fressin}, {Ge}, {Henning},
  {Holman}, {Howard}, {Ida}, {Jenkins}, {Jernigan}, {Johnson}, {Kaltenegger},
  {Kawai}, {Kjeldsen}, {Laughlin}, {Levine}, {Lin}, {Lissauer}, {MacQueen},
  {Marcy}, {McCullough}, {Morton}, {Narita}, {Paegert}, {Palle}, {Pepe},
  {Pepper}, {Quirrenbach}, {Rinehart}, {Sasselov}, {Sato}, {Seager},
  {Sozzetti}, {Stassun}, {Sullivan}, {Szentgyorgyi}, {Torres}, {Udry}, \&
  {Villasenor}}]{2015JATIS...1a4003R}
{Ricker}, G.~R., {Winn}, J.~N., {Vanderspek}, R., {et~al.} 2015, Journal of
  Astronomical Telescopes, Instruments, and Systems, 1, 014003

\bibitem[{{Salaris} {et~al.}(2022){Salaris}, {Cassisi}, {Pietrinferni}, \&
  {Hidalgo}}]{2022MNRAS.509.5197S}
{Salaris}, M., {Cassisi}, S., {Pietrinferni}, A., \& {Hidalgo}, S. 2022,
  \mnras, 509, 5197

\bibitem[{{Serenelli} {et~al.}(2019){Serenelli}, {Rohrmann}, \&
  {Fukugita}}]{2019A&A...623A.177S}
{Serenelli}, A., {Rohrmann}, R.~D., \& {Fukugita}, M. 2019, \aap, 623, A177

\bibitem[{Steinmetz {et~al.}(2020{\natexlab{a}})Steinmetz, Guiglion, McMillan,
  Matijevič, Enke, Kordopatis, Zwitter, Valentini, Chiappini, Casagrande,
  Wojno, Anguiano, Bienaymé, Bijaoui, Binney, Burton, Cass, de~Laverny,
  Fiegert, Freeman, Fulbright, Gibson, Gilmore, Grebel, Helmi, Kunder, Munari,
  Navarro, Parker, Ruchti, Recio-Blanco, Reid, Seabroke, Siviero, Siebert,
  Stupar, Watson, Williams, Wyse, Anders, Antoja, Birko, Bland-Hawthorn,
  Bossini, García, Carrillo, Chaplin, Elsworth, Famaey, Gerhard, Jofre, Just,
  Mathur, Miglio, Minchev, Monari, Mosser, Ritter, Rodrigues, Scholz, Sharma,
  Sysoliatina, \& collaboration)}]{Steinmetz_2020II}
Steinmetz, M., Guiglion, G., McMillan, P.~J., {et~al.} 2020{\natexlab{a}}, The
  Astronomical Journal, 160, 83

\bibitem[{Steinmetz {et~al.}(2020{\natexlab{b}})Steinmetz, Matijevič, Enke,
  Zwitter, Guiglion, McMillan, Kordopatis, Valentini, Chiappini, Casagrande,
  Wojno, Anguiano, Bienaymé, Bijaoui, Binney, Burton, Cass, de~Laverny,
  Fiegert, Freeman, Fulbright, Gibson, Gilmore, Grebel, Helmi, Kunder, Munari,
  Navarro, Parker, Ruchti, Recio-Blanco, Reid, Seabroke, Siviero, Siebert,
  Stupar, Watson, Williams, Wyse, Anders, Antoja, Birko, Bland-Hawthorn,
  Bossini, García, Carrillo, Chaplin, Elsworth, Famaey, Gerhard, Jofre, Just,
  Mathur, Miglio, Minchev, Monari, Mosser, Ritter, Rodrigues, Scholz, Sharma,
  Sysoliatina, \& collaboration)}]{Steinmetz_2020}
Steinmetz, M., Matijevič, G., Enke, H., {et~al.} 2020{\natexlab{b}}, The
  Astronomical Journal, 160, 82

\bibitem[{{Torres} {et~al.}(2022){Torres}, {Canals}, {Jim{\'e}nez-Esteban},
  {Rebassa-Mansergas}, \& {Solano}}]{2022MNRAS.511.5462T}
{Torres}, S., {Canals}, P., {Jim{\'e}nez-Esteban}, F.~M., {Rebassa-Mansergas},
  A., \& {Solano}, E. 2022, \mnras, 511, 5462

\bibitem[{{Torres} {et~al.}(2019){Torres}, {Cantero}, {Rebassa-Mansergas},
  {Skorobogatov}, {Jim{\'e}nez-Esteban}, \& {Solano}}]{Torres2019}
{Torres}, S., {Cantero}, C., {Rebassa-Mansergas}, A., {et~al.} 2019, \mnras,
  485, 5573

\bibitem[{{Torres} \& {Garc{\'{\i}}a-Berro}(2016)}]{Torres2016}
{Torres}, S. \& {Garc{\'{\i}}a-Berro}, E. 2016, \aap, 588, A35

\bibitem[{{Torres} {et~al.}(2015){Torres}, {Garc{\'{\i}}a-Berro}, {Althaus}, \&
  {Camisassa}}]{Torres2015}
{Torres}, S., {Garc{\'{\i}}a-Berro}, E., {Althaus}, L.~G., \& {Camisassa},
  M.~E. 2015, \aap, 581, A90

\bibitem[{{Torres} {et~al.}(2018){Torres}, {Garc{\'{\i}}a-Berro}, {Cojocaru},
  \& {Calamida}}]{Torres2018}
{Torres}, S., {Garc{\'{\i}}a-Berro}, E., {Cojocaru}, R., \& {Calamida}, A.
  2018, \mnras

\bibitem[{{Torres} {et~al.}(2005){Torres}, {Garc{\'{\i}}a-Berro}, {Isern}, \&
  {Figueras}}]{Torres2005}
{Torres}, S., {Garc{\'{\i}}a-Berro}, E., {Isern}, J., \& {Figueras}, F. 2005,
  \mnras, 360, 1381

\bibitem[{{Torres} {et~al.}(2021){Torres}, {Rebassa-Mansergas}, {Camisassa}, \&
  {Raddi}}]{Torres2021}
{Torres}, S., {Rebassa-Mansergas}, A., {Camisassa}, M.~E., \& {Raddi}, R. 2021,
  \mnras, 502, 1753

\bibitem[{{Tremblay} {et~al.}(2019){Tremblay}, {Fontaine}, {Fusillo}, {Dunlap},
  {G{\"a}nsicke}, {Hollands}, {Hermes}, {Marsh}, {Cukanovaite}, \&
  {Cunningham}}]{2019Natur.565..202T}
{Tremblay}, P.-E., {Fontaine}, G., {Fusillo}, N.~P.~G., {et~al.} 2019, \nat,
  565, 202

\bibitem[{{Winget} \& {Kepler}(2008)}]{2008ARA&A..46..157W}
{Winget}, D.~E. \& {Kepler}, S.~O. 2008, \araa, 46, 157

\bibitem[{{York} {et~al.}(2000){York}, {Adelman}, {Anderson}, {Anderson},
  {Annis}, {Bahcall}, {Bakken}, {Barkhouser}, {Bastian}, {Berman}, {Boroski},
  {Bracker}, {Briegel}, {Briggs}, {Brinkmann}, {Brunner}, {Burles}, {Carey},
  {Carr}, {Castander}, {Chen}, {Colestock}, {Connolly}, {Crocker}, {Csabai},
  {Czarapata}, {Davis}, {Doi}, {Dombeck}, {Eisenstein}, {Ellman}, {Elms},
  {Evans}, {Fan}, {Federwitz}, {Fiscelli}, {Friedman}, {Frieman}, {Fukugita},
  {Gillespie}, {Gunn}, {Gurbani}, {de Haas}, {Haldeman}, {Harris}, {Hayes},
  {Heckman}, {Hennessy}, {Hindsley}, {Holm}, {Holmgren}, {Huang}, {Hull},
  {Husby}, {Ichikawa}, {Ichikawa}, {Ivezi{\'c}}, {Kent}, {Kim}, {Kinney},
  {Klaene}, {Kleinman}, {Kleinman}, {Knapp}, {Korienek}, {Kron}, {Kunszt},
  {Lamb}, {Lee}, {Leger}, {Limmongkol}, {Lindenmeyer}, {Long}, {Loomis},
  {Loveday}, {Lucinio}, {Lupton}, {MacKinnon}, {Mannery}, {Mantsch}, {Margon},
  {McGehee}, {McKay}, {Meiksin}, {Merelli}, {Monet}, {Munn}, {Narayanan},
  {Nash}, {Neilsen}, {Neswold}, {Newberg}, {Nichol}, {Nicinski}, {Nonino},
  {Okada}, {Okamura}, {Ostriker}, {Owen}, {Pauls}, {Peoples}, {Peterson},
  {Petravick}, {Pier}, {Pope}, {Pordes}, {Prosapio}, {Rechenmacher}, {Quinn},
  {Richards}, {Richmond}, {Rivetta}, {Rockosi}, {Ruthmansdorfer}, {Sandford},
  {Schlegel}, {Schneider}, {Sekiguchi}, {Sergey}, {Shimasaku}, {Siegmund},
  {Smee}, {Smith}, {Snedden}, {Stone}, {Stoughton}, {Strauss}, {Stubbs},
  {SubbaRao}, {Szalay}, {Szapudi}, {Szokoly}, {Thakar}, {Tremonti}, {Tucker},
  {Uomoto}, {Vanden Berk}, {Vogeley}, {Waddell}, {Wang}, {Watanabe},
  {Weinberg}, {Yanny}, {Yasuda}, \& {SDSS Collaboration}}]{2000AJ....120.1579Y}
{York}, D.~G., {Adelman}, J., {Anderson}, Jr., J.~E., {et~al.} 2000, \aj, 120,
  1579

\end{thebibliography}



\end{document}